\documentclass[twocolumn,11pt]{infocom}

\usepackage{graphicx}
\usepackage{amsmath}

\begin{document}

\title{Analysis of Network Traffic in Switched Ethernet Systems}
\author{Tony Field, Uli Harder \& Peter Harrison 
  \thanks{Department of Computing
    Imperial College of Science, Technology and Medicine
    Huxley Building, 180 Queen's Gate, London SW7 2BZ, UK, Tel: +44
    (0)20 7594 8285  Fax: +44 (0)20 7581 8024, e-mail:
    \texttt{ajf@doc.ic.ac.uk}, \texttt{uh@doc.ic.ac.uk},
    \texttt{pgh@doc.ic.ac.uk}   
    }
  }

\maketitle

\begin{abstract}
A 100 Mbps Ethernet link between a college campus and  the outside
world was monitored 
with a
dedicated PC and the measured data analysed for its statistical
properties. Similar measurements were taken at an internal node of the
network. The networks in both cases are a full-duplex switched Ethernet.
Inter-event interval histograms and power spectra of the
throughput aggregated for 10ms bins were used to analyse the measured
traffic. For most investigated cases both methods reveal that the
traffic behaves according to a power law. The results will be used in
later studies to parameterise models for network traffic. 

\end{abstract}
\section{Introduction}
\label{sec:introduction}
The rapid growth of Internet technologies has created an urgent need for
predictive models of performance.  To this end, it has become
imperative to
obtain a good abstraction of the diverse types of network traffic: partly
in
its own right so that usage patterns can be studied, but mainly to provide
input
to performance models.  

The work described in this paper is part of an ongoing research project 
which seeks to develop accurate performance models of 
large-scale, high-performance IP networks comprising
hundreds or thousands of switched Ethernet routers. 
Networks of this scale are becoming commonplace
and seem likely to grow significantly in number (and probably also
size) in the foreseeable future.  Understanding the performance of
these networks is 
of increasing importance, particularly 
given the trend for guaranteed
quality of service in private networks and between internet service providers
(ISP).  How, 
for example, is service quality affected by a failure in a router or link
and how can a network be engineered to maintain service guarantees in the
presence of failures?
Future proposals for Internet charging, e.g.~\cite{Charging},
 also raise interesting new
challenges that cannot be properly addressed without some model of
network performance.

This paper focuses on one aspect of this exercise, namely that of
monitoring with a view to modelling. We focus on IP traffic which has
been measured by monitoring 
a real network using a high-performance trace capture facility capable of 
monitoring network links up to 100Mbps. 
Fitting the measured IP traffic is complicated by the fact that the
data typically is 
correlated.  IP traffic streams cannot therefore be described accurately by
pure Poisson models \cite{LTWW,ENW}, with the unfortunate consequence
that well-known analytic models are no longer adequate to predict
and analyse performance. Most realistic models put forward turn
out to be mathematically and/or numerically intractable for analytic
solution.
The Markov Modulated Poisson Process (MMPP)\cite{Cookbook}, however,
does have non-zero 
autocorrelation between inter-arrival times and can often be used to
describe effectively the type of traffic observed in networks,
sometimes seen as self-similar.  However, we consider 
self-similarity  synonymously with a time series having   a
heavy-tail distribution (typically polynomial) or shows time-scale
invariance. Self-similarity is defined rigorously in terms of the
latter property.

The paper makes the following contributions:
\begin{itemize}
\item We describe a  monitoring scheme using conventional hardware and
software that is capable of capturing 
a trace of switched Ethernet traffic on links of up
to 100Mbps.
\item We present results of significant behavioural observations made
  using the monitoring scheme at 
different parts of a
university department network.
\item We show findings of preliminary statistical analyses of  
 traces made for a range of different
traffic types at several points in that network
at different times. For some traffic types we find $1/f$ noise, whilst
others appear to be uncorrelated. 
\end{itemize}
The rest of the paper is organised as follows.  In section~\ref{net} the
network under study is described along with the monitoring techniques used
to
collect traffic data.  Section~\ref{sec:stats} describes the
statistical methods used to abstract the critical features of the traffic
in a
quantitative way.
Sections~\ref{bd}, ~\ref{node} and ~\ref{mmpp} present the raw
numerical data and 
the results of the statistical analysis.  The paper concludes with a discussion
of future research plans and directions. Ultimately we would like
incorporate our findings appropriately in queuing models.

\section{Network architecture and monitoring} \label{net}
\begin{figure}[tbp]
\hfill
\begin{center}
\fbox{\includegraphics[height=5.5cm]{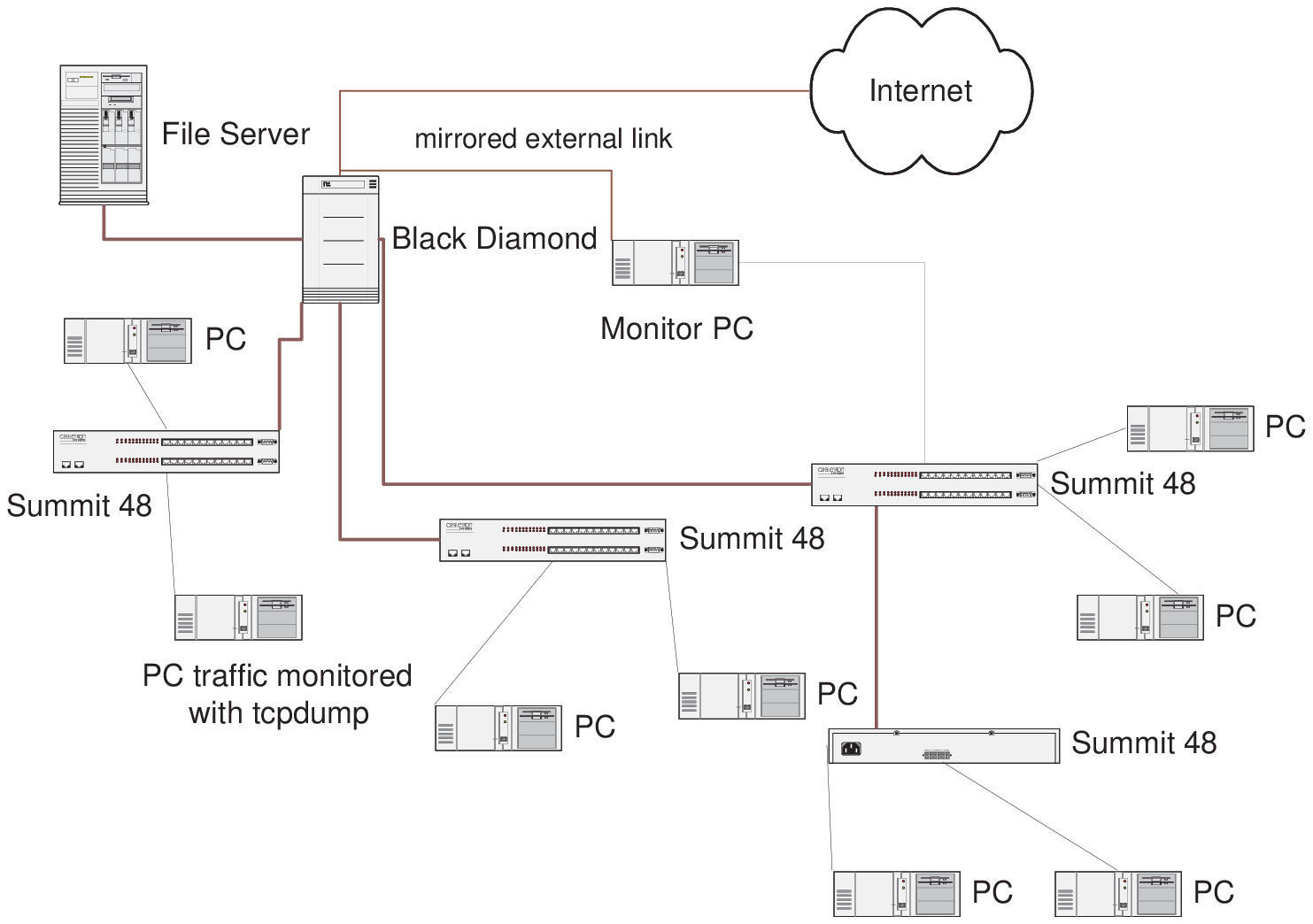}}
\end{center}
\caption{A sketch of the departmental Ethernet. The root switch is the
  Black Diamond. Smaller switches hang off it connected via 1Gbps
  links. If possible there is at most one hop between an end-node
  and the root switch. However, physical constraints make this
  impossible in some cases. The PC monitoring the outside network
  connection has 2 network interfaces. 100 Mbps connections are
  depicted as thinner lines. }
\label{network-sketch}
\end{figure}

\begin{figure}[tbp]
\hfill
\begin{center}
\fbox{\includegraphics[height=9cm]{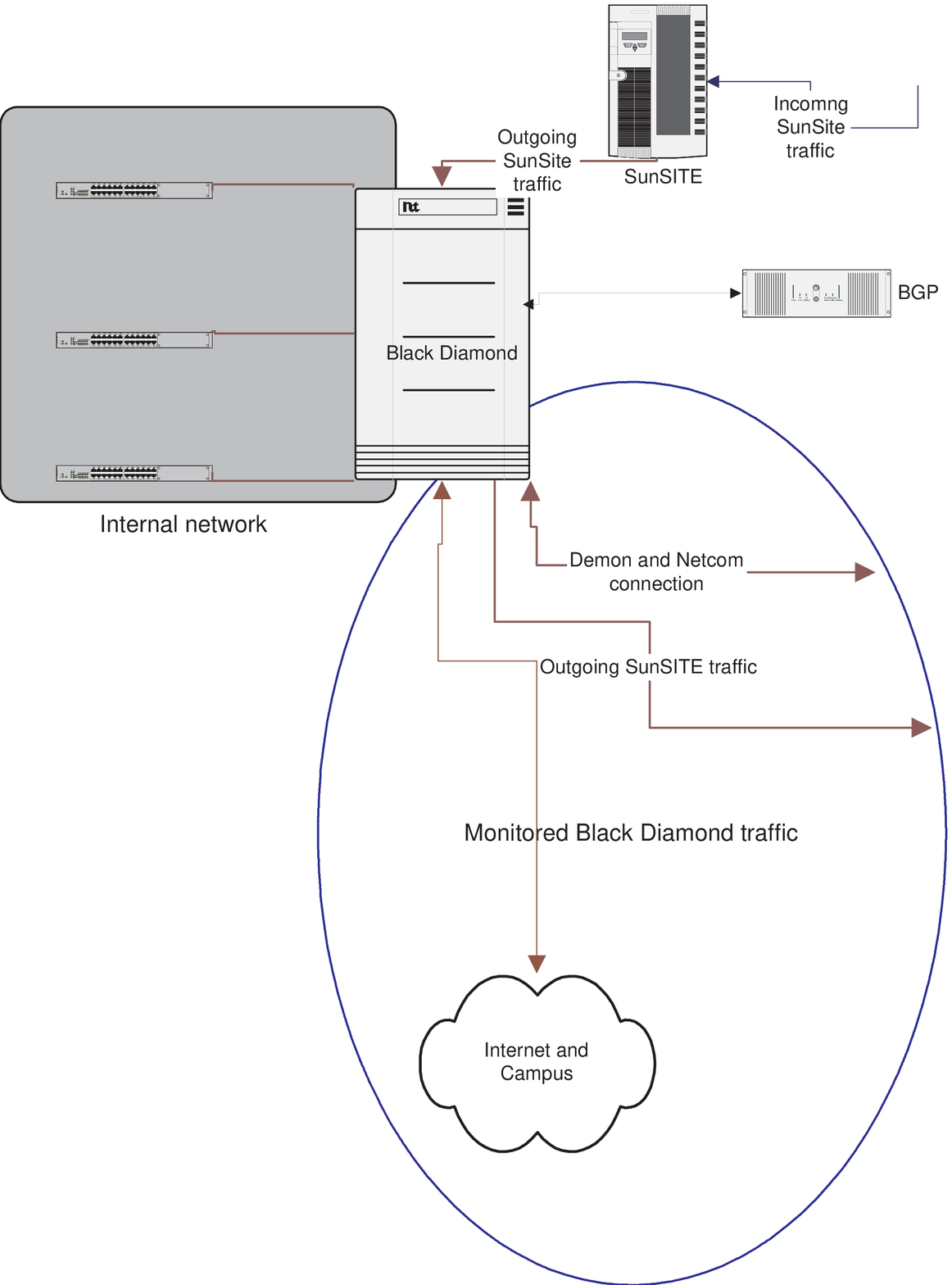}}
\end{center}
\caption{\small Overview of the network we have monitored. Note that the
  activity of the core router due to the internal network is not
  recorded, neither is the incoming SunSITE traffic.}
\label{network-overview}
\end{figure}

The network monitoring work was carried out on the Department of Computing's
network at Imperial College in London.  This comprises a network of full-duplex
switched Ethernet  
hubs. Each node (a workstation or PC) is connected 
via  a 100Mbps link  to either a
24-way or 48-way switched Ethernet router 
(Extreme `Summit-24' or `Summit-48' routers).
These are connected in turn to a central hub (a 'Black Diamond') \footnote{In the
following we will also refer to this part of the network synonymously
as the central 
hub or
core router to distinguish it from the smaller hubs. The distinction
between hubs, bridges and routers is slightly blurred as the
department is connected to the campus via an Ethernet and  the
smaller hubs are switched. }
 whose current configuration can support up to 24 1Gbps links
and several 100Mbs links \cite{EN}.
Various
file and CPU servers also hang off the
central hub.
In general every node is only one
hop away from the central hub although there are a few exceptions.  
The sketch in (fig. \ref{network-sketch}) illustrates
the basic topology of the network.
The core router  is used for the internal and external
network traffic. Internally it used when nodes from different switches
exchange data over the network. Externally it is used for the in and
outgoing traffic to nodes in the department (PCs, web-server,
mail-server, news-server etc.) and also for the outgoing traffic of
the SunSITE \cite{sunsite} ftp server. Additionally there are 2
connections to ISPs, Demon and Netcom.\footnote{The 
mirrored ports we monitor
do not include the internal
departmental traffic. The latter has to be excluded due to the
suspected high volume. Also, the internal traffic is handled on a different
level by the Black Diamond as it simply forwards Ethernet frames based
on media access control (MAC) addresses. For the external traffic,
routing tables have to be 
consulted.} This is illustrated in (fig. \ref{network-overview}).

An important distinction between this network and many others
previously studied is the Ethernet
connection regime.  Unlike conventional Ethernet, the switched
Ethernet used in the department is collision-free. A
conventional Ethernet is a single shared 
resource (collision domain) 
which can only be used by one node at a time.
Attached nodes contend for the Ethernet and, once claimed, hold it for the
duration of a data transfer.  Access to the Ethernet is managed by
the  CSMA/CD protocol which essentially implements a collision detection
and back-off algorithm. 

By contrast, in a full-duplex switched Ethernet 
all nodes have a separate network connection for inward- and
outward-bound traffic. Each hub contains a switch,
which is a multiplexer that connects two switch ports.
Contention for a switch output is resolved by queueing, in contrast to
collisions
in a conventional Ethernet.  Essentially, the collision domain has been
broken down into a number of separate collision domains between each node and
its 
switch. Routing decisions may be taken directly
by the switch or may be referred to a table look-up prior to switching.
At this stage we are not interested in modelling
the activities within a router so the distinction is not
important. For a good introduction to the details of Ethernet
technology see \cite{Spur}.
\subsection{Monitoring}
To monitor a conventional shared Ethernet one simply needs to switch
one network 
interface card (NIC) into {\em promiscuous} mode in which the interface
``listens'' to all 
packets sent along the carrier. This option is not
available for switched networks as each node only sees the 
data that is destined for itself. 
This benefits security,
but makes it necessary to use different methods for monitoring.
One option is the simple network monitoring protocol \textsc{
  Snmp}~\cite{WS} but this suffers from
lax security which makes it relatively easy to attack, essentially by
remotely reprogramming the switches.

SNMP has therefore been disabled and the monitoring
problem has been resolved by arranging for
the traffic on one port to be {\em mirrored} to another.  Data is
captured from this mirrored port using \verb|tcpdump| ~\cite{tcpdump}
which generates 
a summary of each Ethernet frame passing through the port in either
direction.\footnote{Only the first 150 bytes of each packet are captured. This
  is sufficient to gather information about source, destination, size
  and type of the frame.}
The 100Mbps links can be trace-captured in this way using a PC of modest
power\footnote{A Celeron 400MHz, 128 Mb RAM with four fast SCSI disks, in our
case running Linux}.  
Here, we report results for traffic to and from the
outside world through the Black Diamond router 
and a single  PC using the same (mirroring) 
technique (100Mbps links in both cases).  This enables the individual
traffic streams to and from 
individual nodes to be compared with the aggregate traffic seen
between switches. 

Analysis of data from \verb|tcpdump| reveals information about higher level
protocols that use Ethernet frames to transmit their packets. 
This is mainly IP (used by TCP and UDP). \verb|tcpdump| actually
reports on all Ethernet frames that pass the NIC. So the name
is slightly misleading. We have used the program to get 
\begin{itemize}
\item A timestamp indicating when the kernel has ``seen'' the packet
\item Source and destination IP address and port number
\item The size of the frame (only the user data is reported, the headers for
  various protocol layers have to be added on to recover the actual
  size of the frame) 
\item The traffic type (tcp, udp, icmp, etc.) 
\end{itemize}
for every frame that is transmitted.

\subsection{Measurement errors}
\label{error}

We gained some idea of the size of the error in our measurements by
looking at the reported inter-arrival times. Given the time stamp and
size of the previous packet one can determine the time the Ethernet
has been busy and therefore when the next arrival can possibly have
happened. We noticed that the reported times were up to 100 $\mu$sec too
early in a number of cases. This is due to several factors
\begin{itemize}
\item The network interface card introduces errors by buffering
  data unreliably
\item \verb|tcpdump| runs as a user-level process and hence can simply
  miss CPU cycles if the machine is under a heavy load
\item PC-based hardware is less reliable than, for instance, SPARC
  hardware when it comes to small timings \cite{TCPIP-ill}. 
\end{itemize}
The only way to get better  timings is to spend more money on
equipment. We assume that our measurement error is around 50
$\mu$sec. In general \verb|tcpdump| seems to be a fairly reliable
program as has been pointed out by \cite{BT}. Another reason for the
seemingly implausible inter-arrival times may be the fact that we
monitor a full-duplex connection which is able to send and receive
simultaneously. However, the kernel of the monitoring OS can only deal
with one packet at a time and will hence introduce errors if sending
and receiving happens closely together. 

\section{Traffic Analysis}
Two of the first investigations of the statistical nature of network traffic
were \cite{LTWW,ENW}. The authors found evidence that the observed
traffic did not conform with  the assumption  that inter-arrival
times of frames were uncorrelated and could be modelled by Poisson
arrivals. They used methods that had been developed earlier by Hurst
who was investigating the ``ideal'' size of reservoirs (a good summary
of Hurst's work  is given in \cite{BM,JF}). In particular Hurst
introduced the rescaled range statistic $R/S$ which gives an idea of the
self-similarity or long-range dependence of a time series. Many other
statistics, which are all proved or conjectured to be related to the Hurst
parameter, have since been introduced. A good review of the estimators and their
relationships can be found in \cite{PK,TTW,TLFHFT}. The next section is
based on the material found in those papers. In this investigation we
will use the power spectrum to analyse the correlation of the
monitored data and
inter-event interval histograms to analyse the inter arrival time
distribution. 

Other methods to investigate the correlation of the data are the
rescaled range statistic, the log-variance plot, detrended
fluctuation analysis, the Fano factor and the Allan factor. For
details of these methods we refer to \cite{BM,JF,PK,TTW,TLFHFT,JB}.
\subsection{Statistical Methods}
\label{sec:stats}
\subsubsection{Inter-event interval histograms}
The time series $X(t)$ resulting from our measurements describe point
processes. One way to characterise the behaviour of such a  process is to
compute the distribution of inter arrival times of the events. 
To do this, we plot the inter-event interval histograms (IIH). Of course,
we need to keep in mind that our inter-arrival times may show significant
inaccuracies if the time between events is less than 50
$\mu$sec (see section \ref{error}). 

The plots we show are all double-logarithmic plots. The bins grow
exponentially in size, i.e. cover intervals of 0 to 2 $\mu$
seconds followed by [2,4) [4,8) , \dots . The y-axis of the plot is
the number of inter arrival times falling into a given bin divided by
the size of 
the bin (to approximate a density function  and make sure that bigger bins do not get a
bigger weight) and also divided by the total number of arrivals (so
that we can compare different observation periods). 

For most plots we find that a large part of the resulting graph can be
fitted to straight line. This implies that there is a power law
behaviour of the probability density function $p(x)$

$$p(x) \propto \beta x^\gamma . $$

This is a characteristic behaviour for heavy tail distributions like
Pareto, L\'evy, Cauchy, Zipf etc. We have not attempted
to fit any particular distribution because we want to investigate in
future work how universal this behaviour is and what its causes are.

Most importantly we can say that none of the histograms would suggest an
exponential distribution similar to the plot for an MMPP in 
(fig. \ref{fig:mmpp}), although this is not clearly contradicted for
some cases. 

Note that the error bars in the histograms have been calculated by
averaging the 
histograms of several different observation periods and correspond
simply to their standard deviation.

\subsubsection{Power Spectrum}
To gather information about the correlation of the point process
observed, one can go down many avenues. 
For a time series of length $n$ the auto-correlation function at lag
$k$ is defined as  
$$
c_k = \frac{1}{\sigma^2(n-k)} \sum (x_i - \hat x) (x_{i+k} - \hat x)
$$
where $\hat x$ is the mean and $\sigma^2$ the variance.
In fact it is usually easier to work with the Fourier transform of the
auto-correlation function. By the Wiener-Khintchine theorem
\cite{WK}, under certain assumptions, this is the same as the power
spectrum (density) of the time series signal \footnote{We should note
  that this is not a very rigorous way of defining the power spectrum,
  as the time series has to fulfil certain criteria for the integral
  to be well defined, for instance.}
$$
S(f) = \lim_{T\to \infty} \frac{1}{4\pi T} \Big| \int_{-T}^T dt X(t)
e^{-i2\pi ft}\Big|^2 .
$$

Since the actual point
process tends to be 
rather sparse it is best to turn the time series into an aggregate
time series of counts. In our investigation we have used 10ms bins for
the aggregation in line with previous research ~\cite{LTWW,ENW}. Again
one is looking for power laws where the power spectrum
$S(f)$ 
behaves like $S(f) \propto 1/f^\alpha,$
where $f$ is the frequency. The exponent $\alpha$ turns
out to be 0 for white noise and 2 for a Brownian motion. From the
relation of the power spectrum to the auto-correlation function it also follows
that an exponent $\alpha$ close to but smaller than 1 corresponds to long time
correlations

If a times series $X(t)$ exhibits scaling laws, i.e. if $X(ta) = g(a)X(t)$
for some  function $g(a)$, it has to exhibit power law
behaviour as $X(t) = b g(t)$ and $g(a) = a^c$ is the only non-trivial
solution to the above 
functional equation. This behaviour is then of course related to
what is known 
as self-similarity. In fact, the exponent $\alpha$ is related to the
Hurst factor. Also, there are many more measures, like
the Fano factor  which can be used to characterised the time
series.

The power spectra were computed using standard methods published in
the {\em Numerical Recipes in C} with overlapping windows \cite{WP}.

\subsection{Measurements taken at the core router}
\label{bd}
\subsubsection{Overall traffic}
\begin{figure}[tbp]
\hfill
\begin{center}
\includegraphics[height=6.5cm,angle=270]{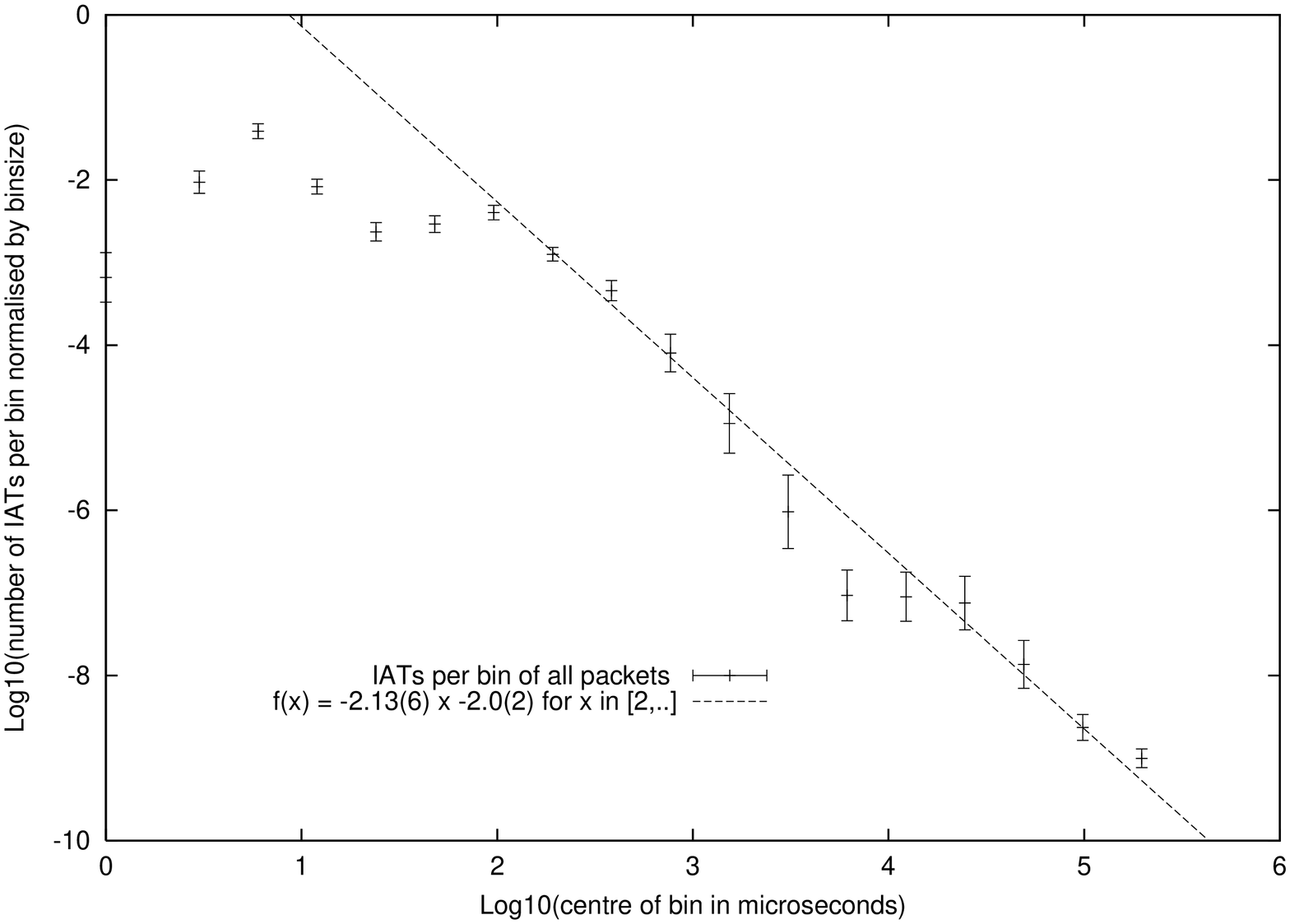}
\includegraphics[height=6.5cm,angle=270]{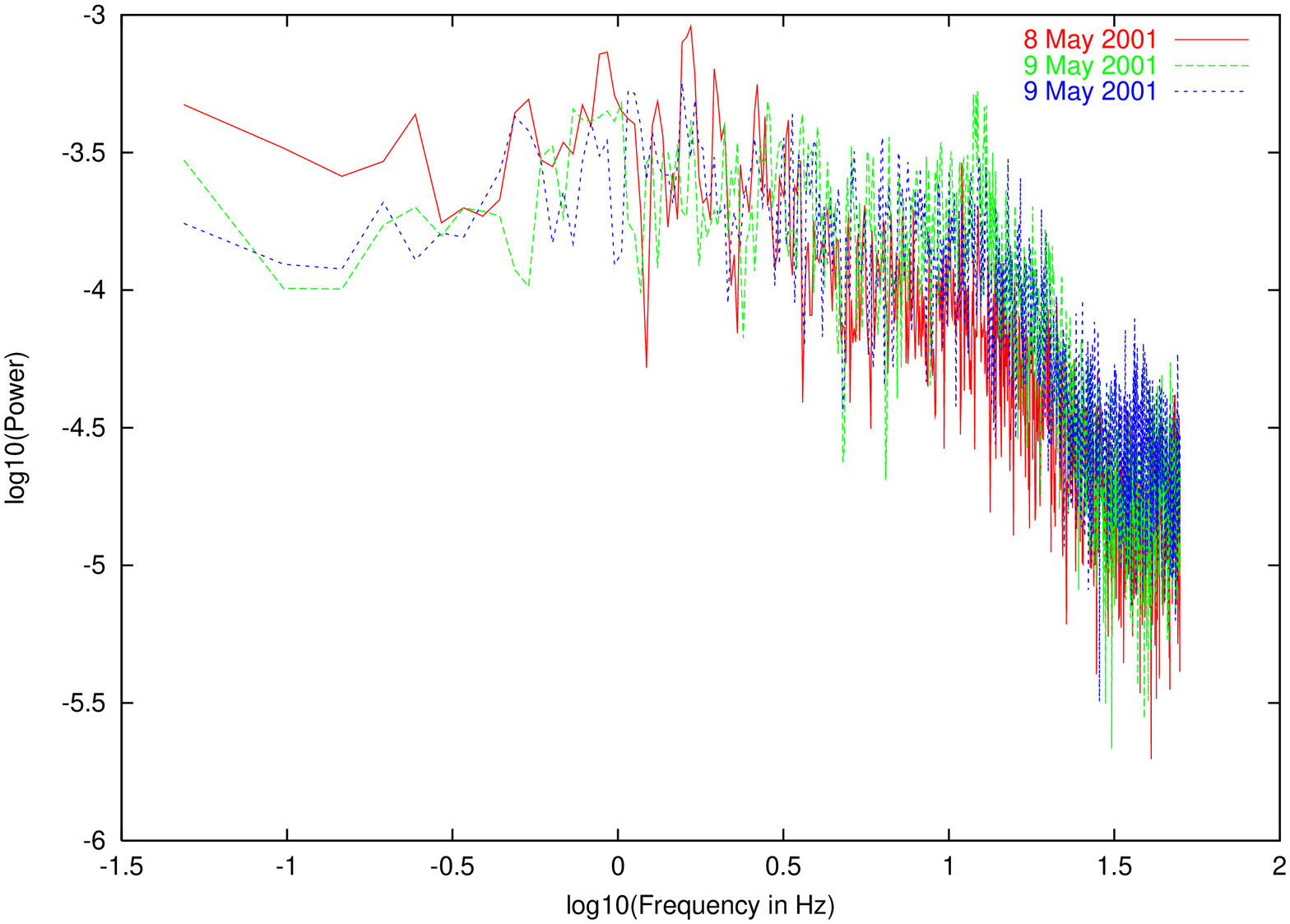}
\end{center}
\caption{\small Plots of Ethernet traffic at the Black Diamond, for
  both in and outgoing traffic combined. The data was collected between 12.30pm
  and 12.35pm on Mondays to Thursdays for 18 days in March, April and
  May. The top graph is the IIH and the bottom graph the power
  spectrum of the utilisation measured in Bytes/sec aggregated for 10ms
  bins. Since the power spectrum does not seem to exhibit power law
  behaviour, we plotted the spectrum for three different days to see
  whether the date first chosen was in some way singular.}
\label{fig:all-bd}
\end{figure}

\begin{figure}[tbp]
\hfill
\begin{center}
\includegraphics[height=6.5cm,angle=270]{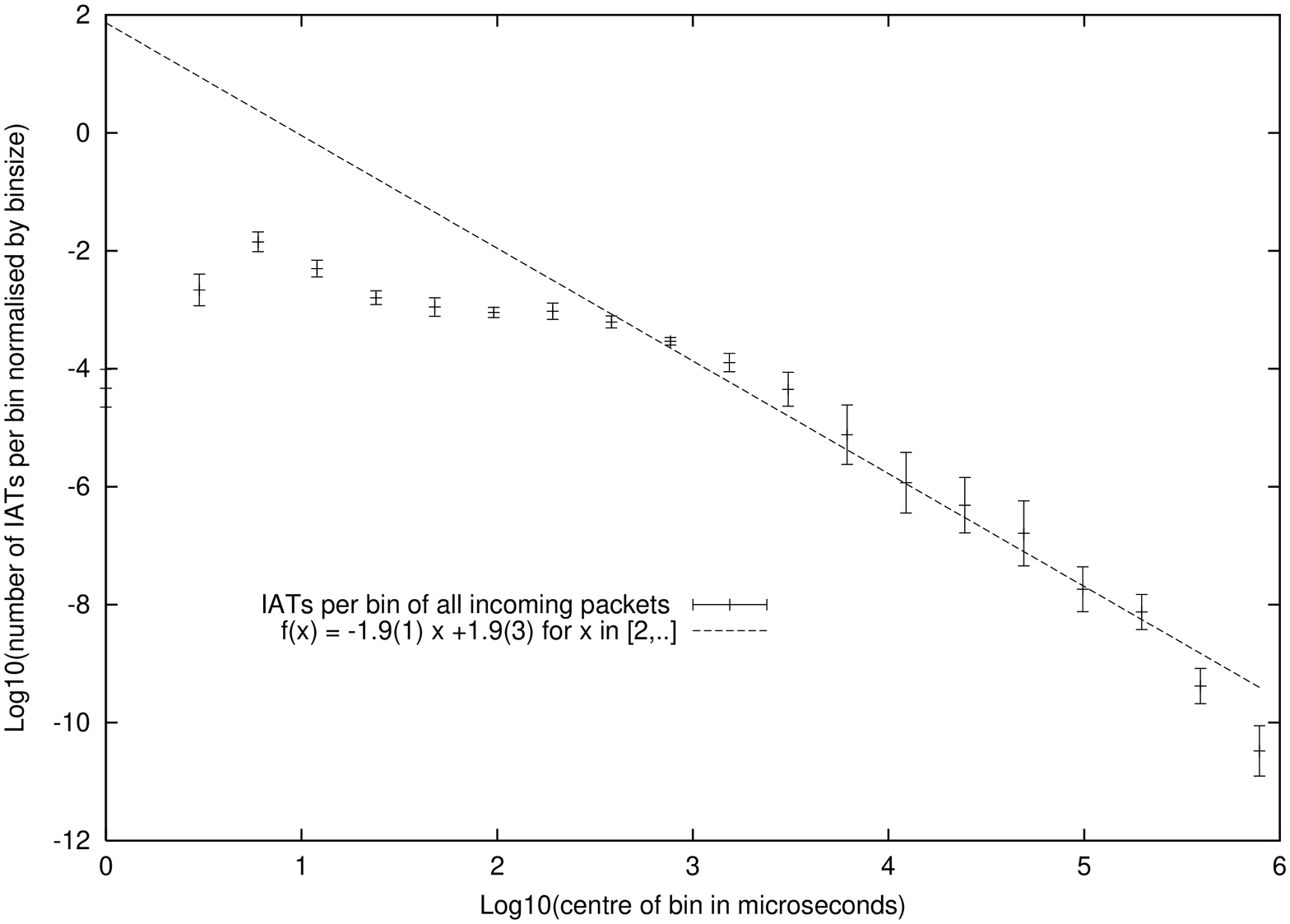}
\includegraphics[height=6.5cm,angle=270]{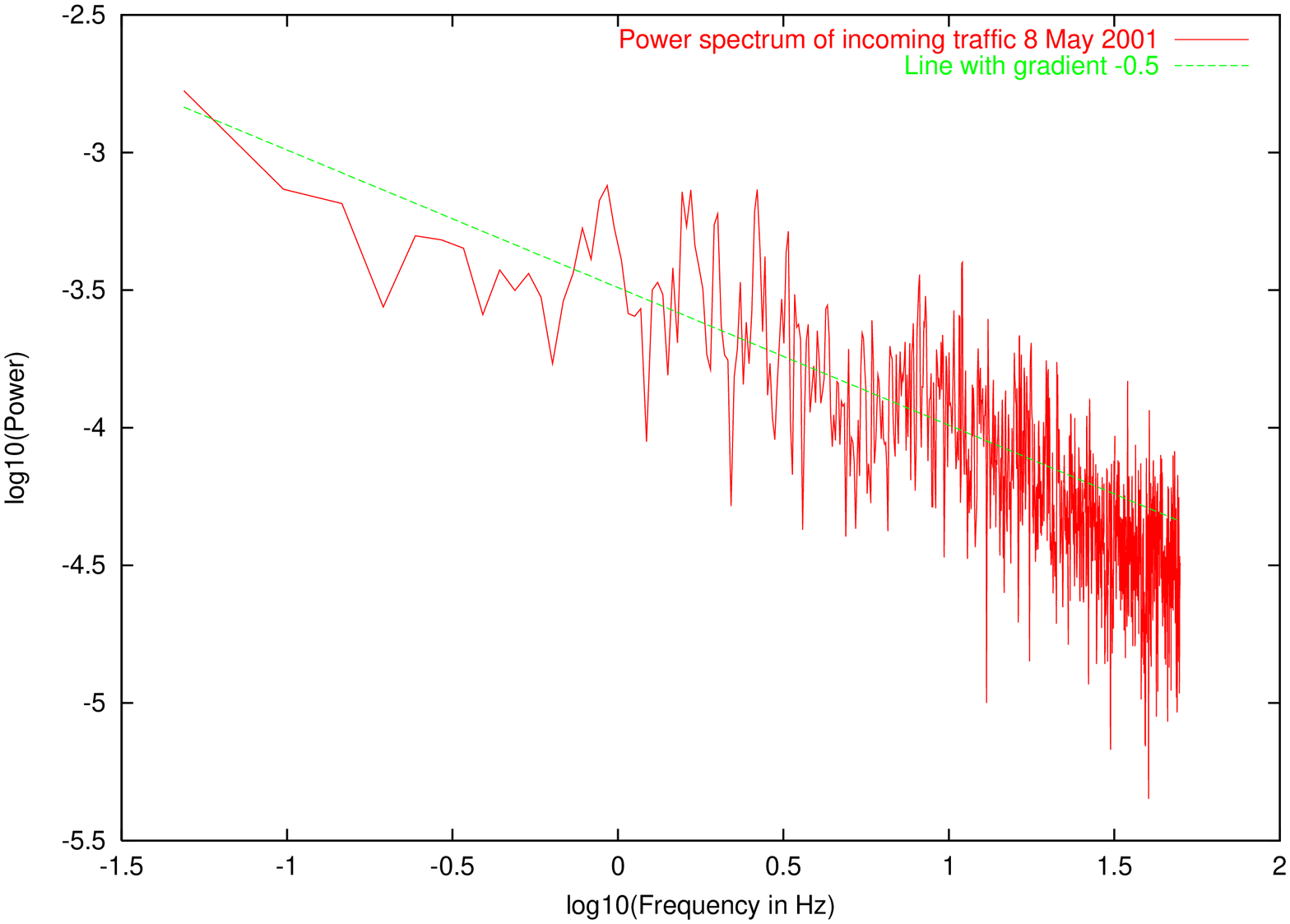}
\end{center}
\caption{\small Plots of Ethernet traffic at the Black Diamond for incoming
  traffic. The plots do not include
  the data that is exchanged with the SunSITE ftp/www server. The data was collected between 12.30pm
  and 12.35pm on Mondays to Thursdays for 18 days in March, April and
  May. The top graph is the IIH and the bottom graph the power
  spectrum of the utilisation measured in Bytes/sec aggregated for 10ms
  bins.}
\label{fig:in-bd}
\end{figure}

\begin{figure}[tbp]
\hfill
\begin{center}
\includegraphics[height=6.5cm,angle=270]{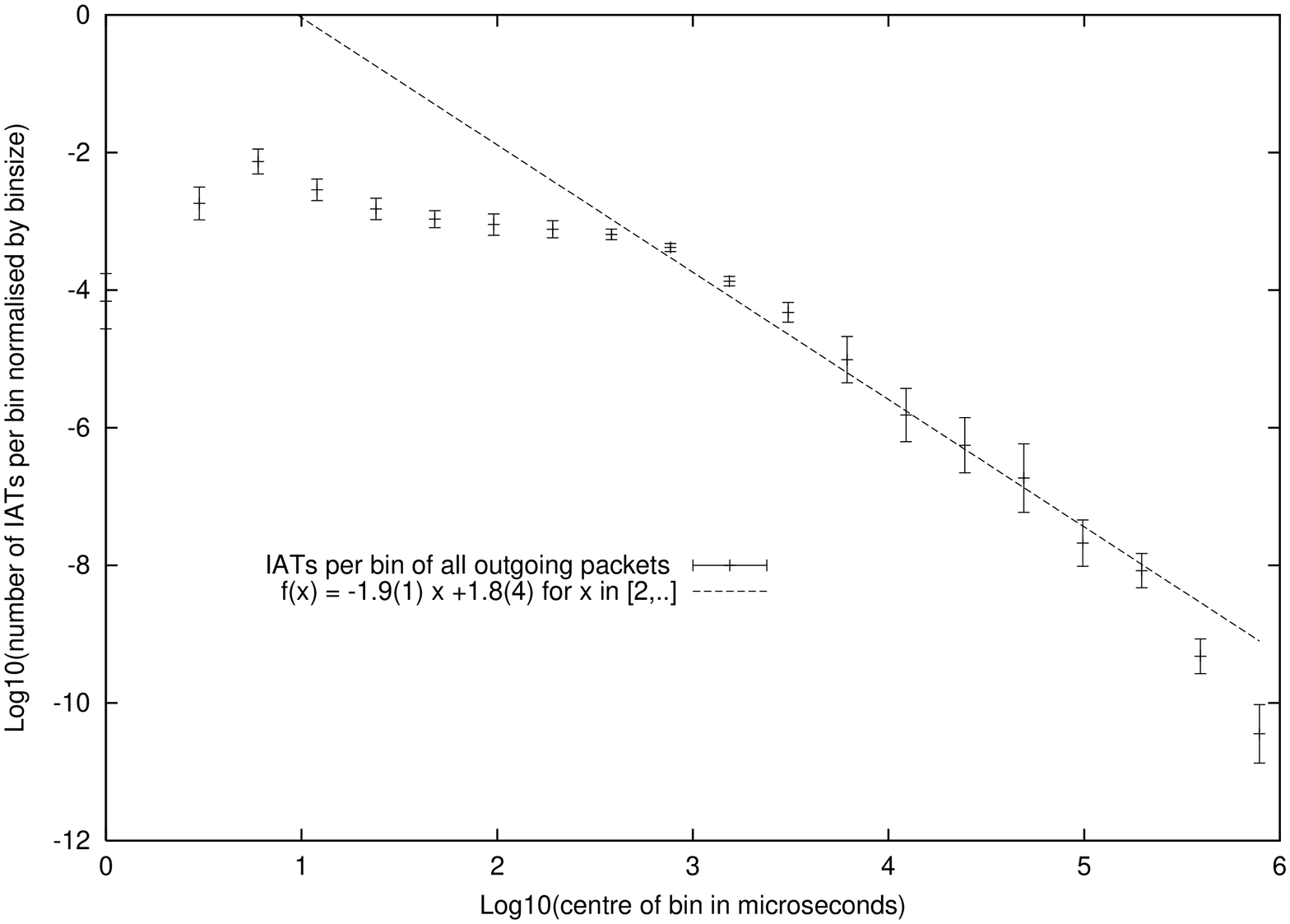}
\includegraphics[height=6.5cm,angle=270]{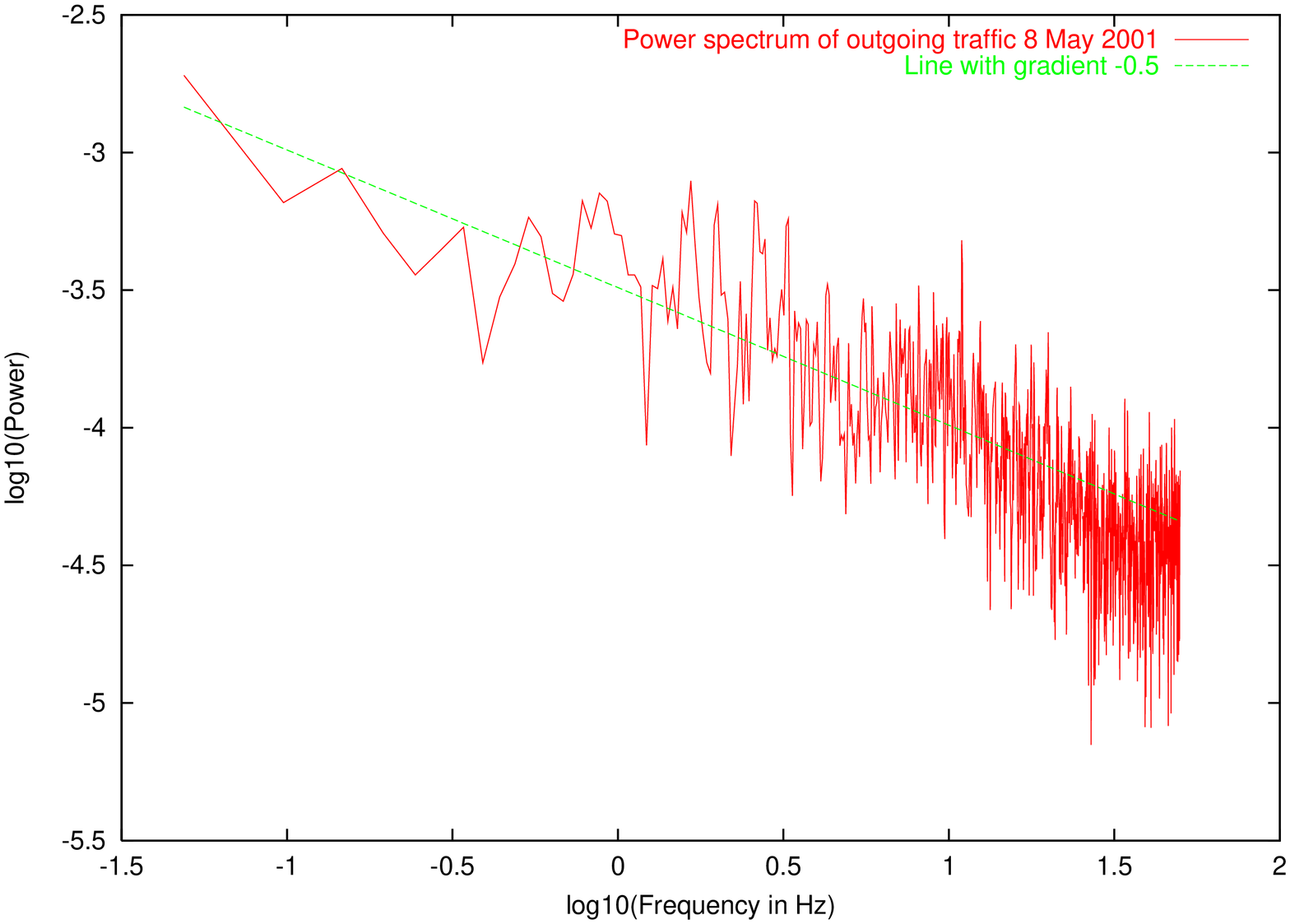}
\end{center}
\caption{\small Plots of Ethernet traffic at the Black Diamond for
  outgoing traffic. The plots do not include
  the data that is exchanged with the SunSITE ftp/www server. The data
  was collected between 12.30pm 
  and 12.35pm on Mondays to Thursdays for 18 days in March, April and
  May. The top graph is the IIH and the bottom graph the power
  spectrum of the utilisation measured in Bytes/sec aggregated for 10ms
  bins.}
\label{fig:out-bd}
\end{figure}

The first set of plots (fig. \ref{fig:all-bd}-\ref{fig:out-bd}) shows
some results from the 
measurements taken 
at the core router\footnote{This is the machine labeled ``Black
  Diamond'' in figures \ref{network-sketch}and \ref{network-overview}}. We
  looked at the traffic that was observed for 18 
days between 12.30 and 12.35. We neglected Fridays, Saturdays and
Sundays as they may well have slightly different characteristics as
users may go home earlier or stay at home in the first place. 

The
traffic is dominated by the \verb|ftp| traffic of {SunSite} which
explains why the in and outbound graphs (fig. \ref{fig:in-bd} and
\ref{fig:out-bd}) for the histograms look very
different to the graph in (fig. \ref{fig:all-bd}), as they do not
include the SunSITE traffic. The graphs
(fig. \ref{fig:all-bd}-\ref{fig:out-bd}) suggest 
that the distribution of the inter arrival 
times might  follow a power law. We suspect that the left-hand side
of the graphs are mainly shaped by measurement errors. However this is
difficult to say for sure without more accurate measurements. It is
interesting to note that the power spectrum of the entire traffic
(fig. \ref{fig:all-bd}) 
seems to fall into two distinct parts and looks very different from
the traffic that does not include the \verb|ftp| traffic
(fig. \ref{fig:all-bd}-\ref{fig:out-bd}). Also, it 
seems difficult to claim that it would follow a power law. 
The plot of the power spectrum actually shows three different days
and since all of them 
show the same behaviour in the power spectrum in
(fig. \ref{fig:all-bd}), this  seems to be not
a singular observation. 

The power spectrum for the incoming traffic suggests a behaviour that can be
well approximated by a power law.
Similarly, the outgoing traffic shows fairly well power law
characteristics that indicate some long-term correlation. In all
graphs there appears to be a strong contribution at the lowest
frequencies. 
Further investigation is necessary  to understand the
power spectrum of the overall traffic better. One
reason for its surprising nature  may be the fact
that we only monitor one half of the traffic (see
fig. \ref{network-overview}). However, if this was the case one would
expect the other power spectrum to have similar features. Another
reason could be the fact that the traffic is mainly \texttt{ftp}
traffic and this application causes the feature.

All histograms plotted so far have shown behaviour for the bins under
0.1 milliseconds that is not consistent with a power law. Due to our
limitation in the accuracy of the timestamp we cannot say whether
this is a feature of the traffic or an artifact of our
measurements. We have neglected the very large bins
for most fits as well because the number of events for these is very
small compared with other bins. 

We can, however, say with some confidence that the network traffic does
indicate the existence of correlation and that the assumption of
exponential inter-arrival time is not in line with our
measurements. Rather we seem to have an inter-arrival time distribution
that follows a power law.

\subsubsection{Web traffic: an example for a single traffic class}

We also looked at traffic that is likely to be caused by web related
use. The way we identified this traffic in our traces was to look for a
server with  port number 80, hence the cautious statement as there
is no obligation to run only web servers on that port. For these
plots we also filtered out packets that were exactly 64 Bytes. The
reason behind this is that these packets are likely to be caused by
opening and  closing TCP connections and also by acknowledgements of data. 

\begin{figure}[tbp]
\hfill
\begin{center}
\includegraphics[height=6.5cm,angle=270]{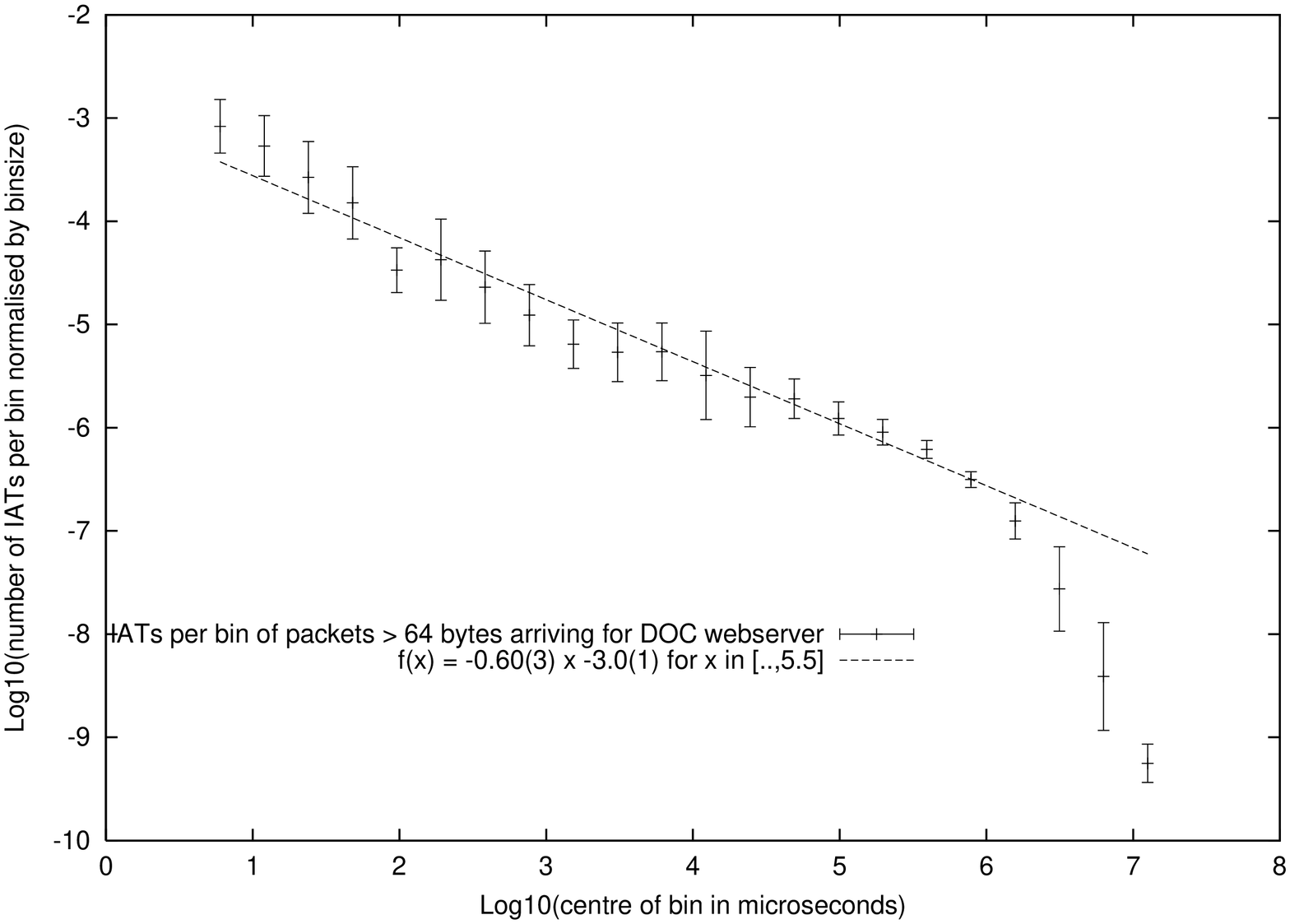}
\includegraphics[height=6.5cm,angle=270]{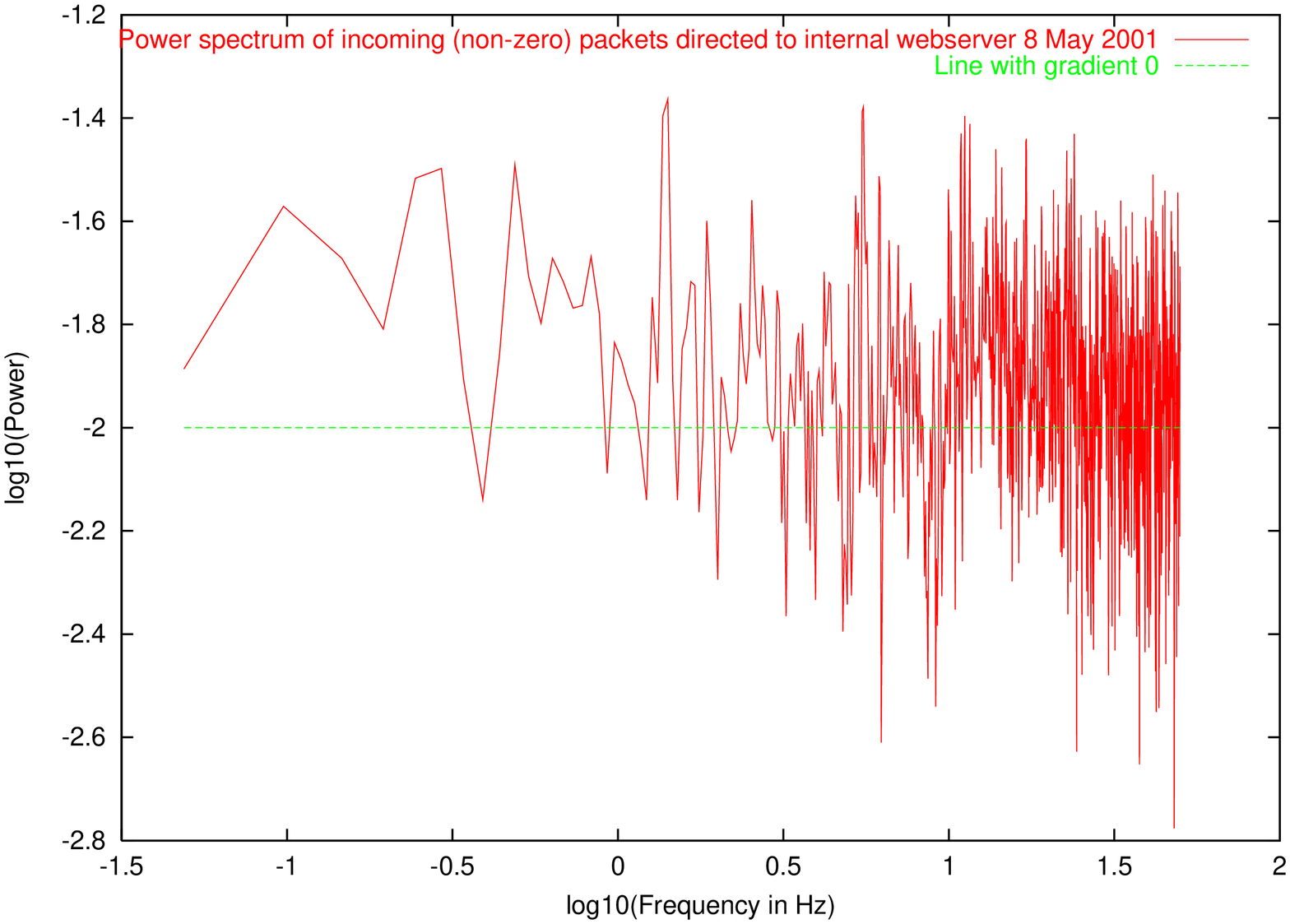}
\end{center}
\caption{\small Probability density function of the inter-arrival times
  for incoming (non-zero) packets for the internal DOC web servers,
   i.e. requests for
  web pages. The power spectrum was taken on the 8 May 2001}
\label{fig:tcp-inin}
\end{figure}

The
first graph (fig. \ref{fig:tcp-inin}) shows the inter-arrival time
distribution of frames that 
come from the outside for a server on port 80, hence (usually) requests to one
of the web servers running in the department. The inter-arrival time
distribution is well 
described by a power law. The power spectrum has  hardly any
gradient at all which means that it may be  nothing but white noise and
therefore has no long-range dependence. Therefore is may be
possible to model this particular stream of traffic as Poisson
arrivals. 

\begin{figure}[tbp]
\hfill
\begin{center}
\includegraphics[height=6.5cm,angle=270]{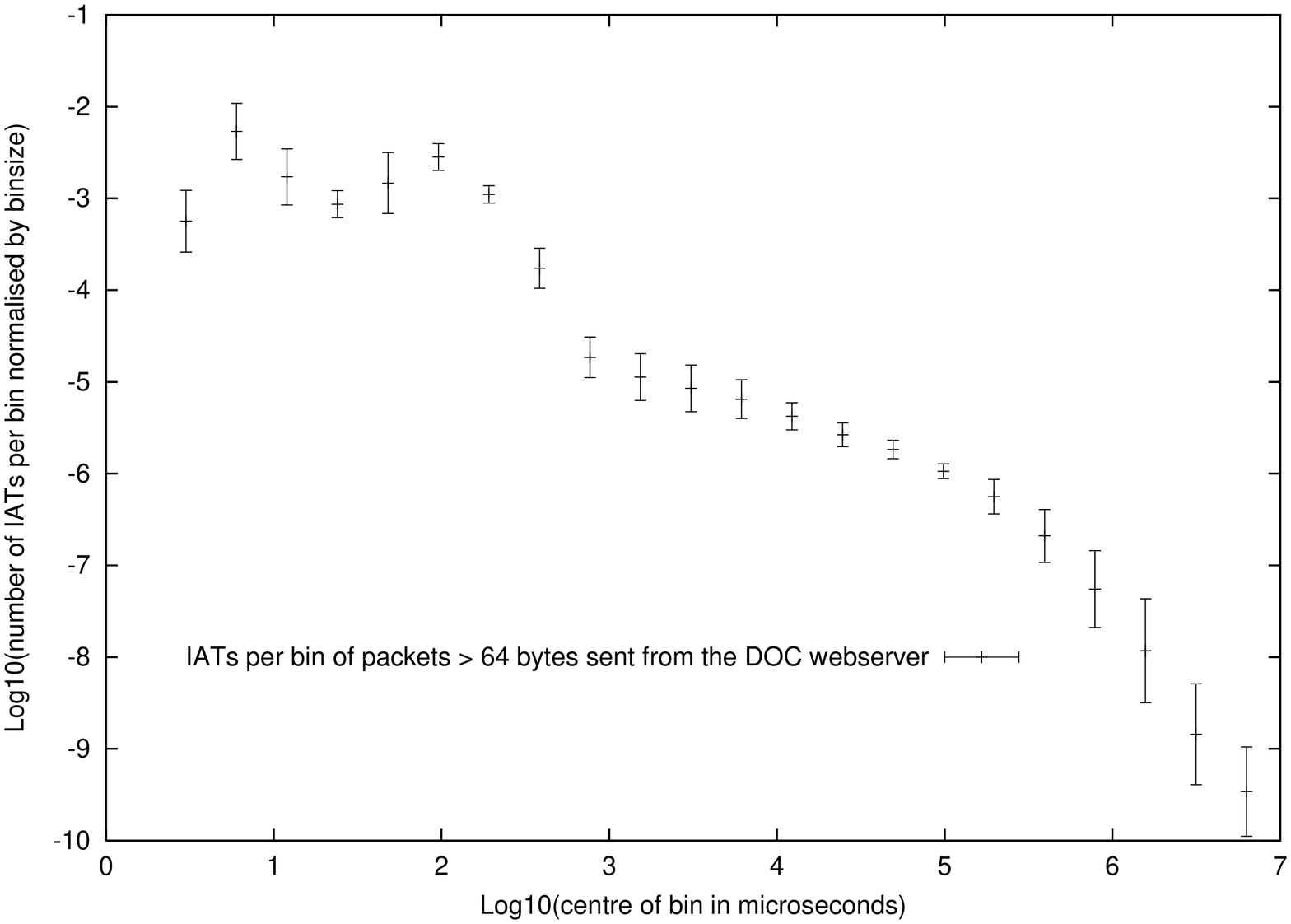}
\includegraphics[height=6.5cm,angle=270]{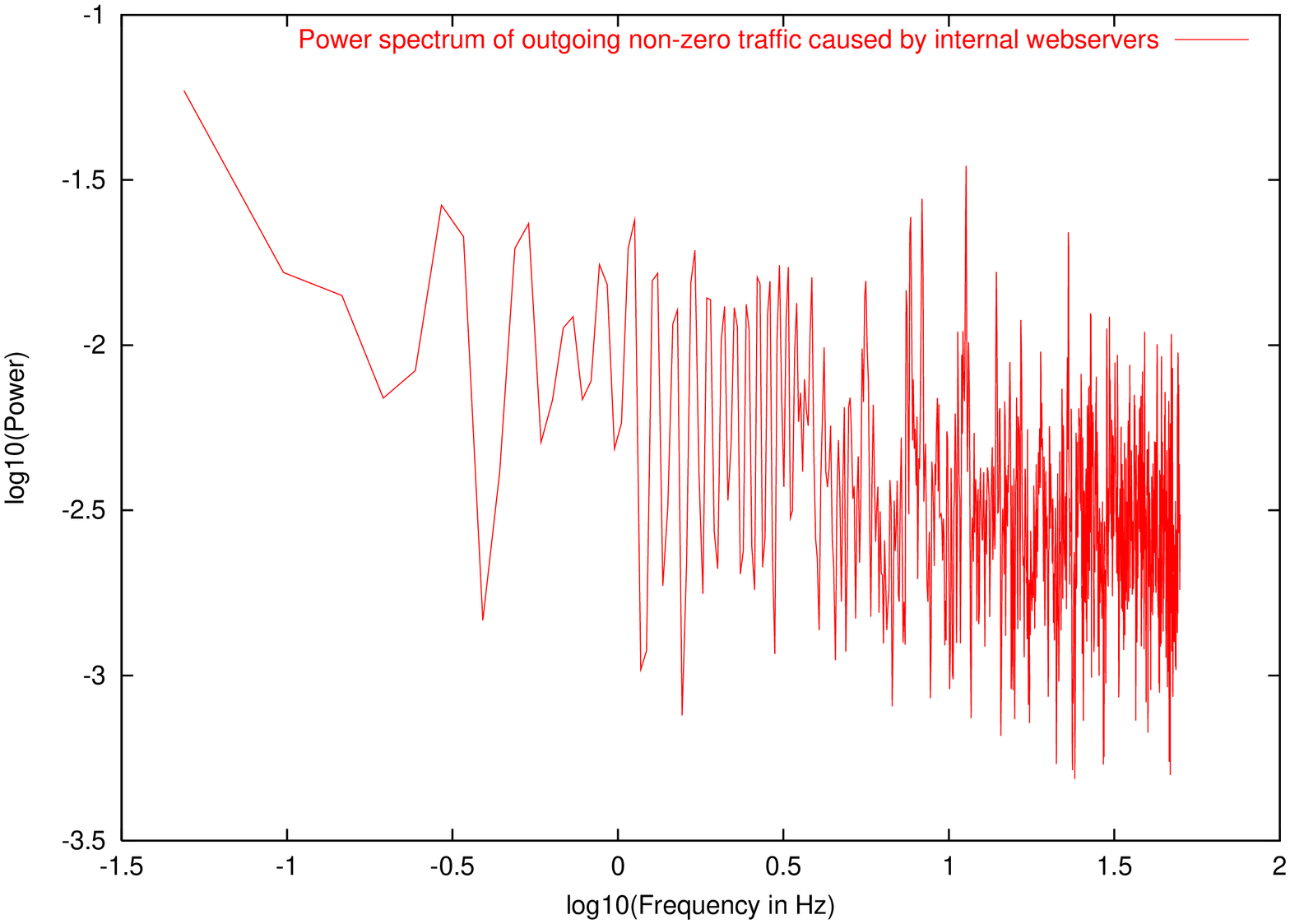}
\end{center}
\caption{\small Probability density function of the inter-arrival times
  for 
  outgoing (non-zero) packets from internal web servers, i.e. replies
  to requests. The power spectrum was taken on the 8 May 2001}
\label{in-out}
\end{figure}

The second set (fig. \ref{in-out}) shows the replies from the internal
web servers. The IIH does not seem to follow a power law in an obvious
way, although one could fit a straight line for the bins between 2.5 and
5 along the x-axis. Partly this plot looks ``worse'' due to a
slightly  different scale of the y-axis compared to the other
plots. The power spectrum shows little correlation. One would expect
the replies to be uncorrelated as the requests already were (see
fig. \ref{fig:tcp-inin}). 

\begin{figure}[tbp]
\hfill
\begin{center}
\includegraphics[height=6.5cm,angle=270]{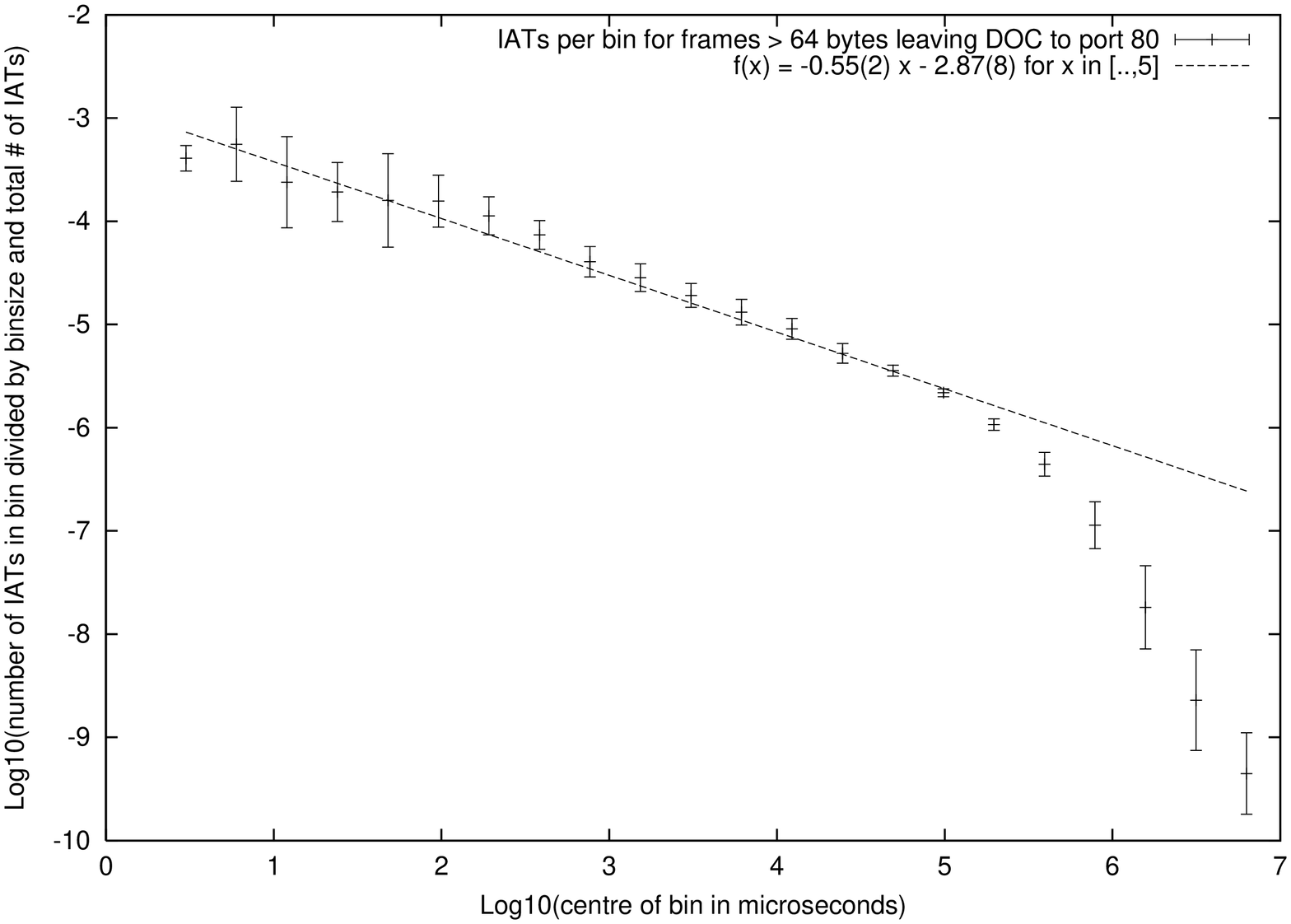}
\includegraphics[height=6.5cm,angle=270]{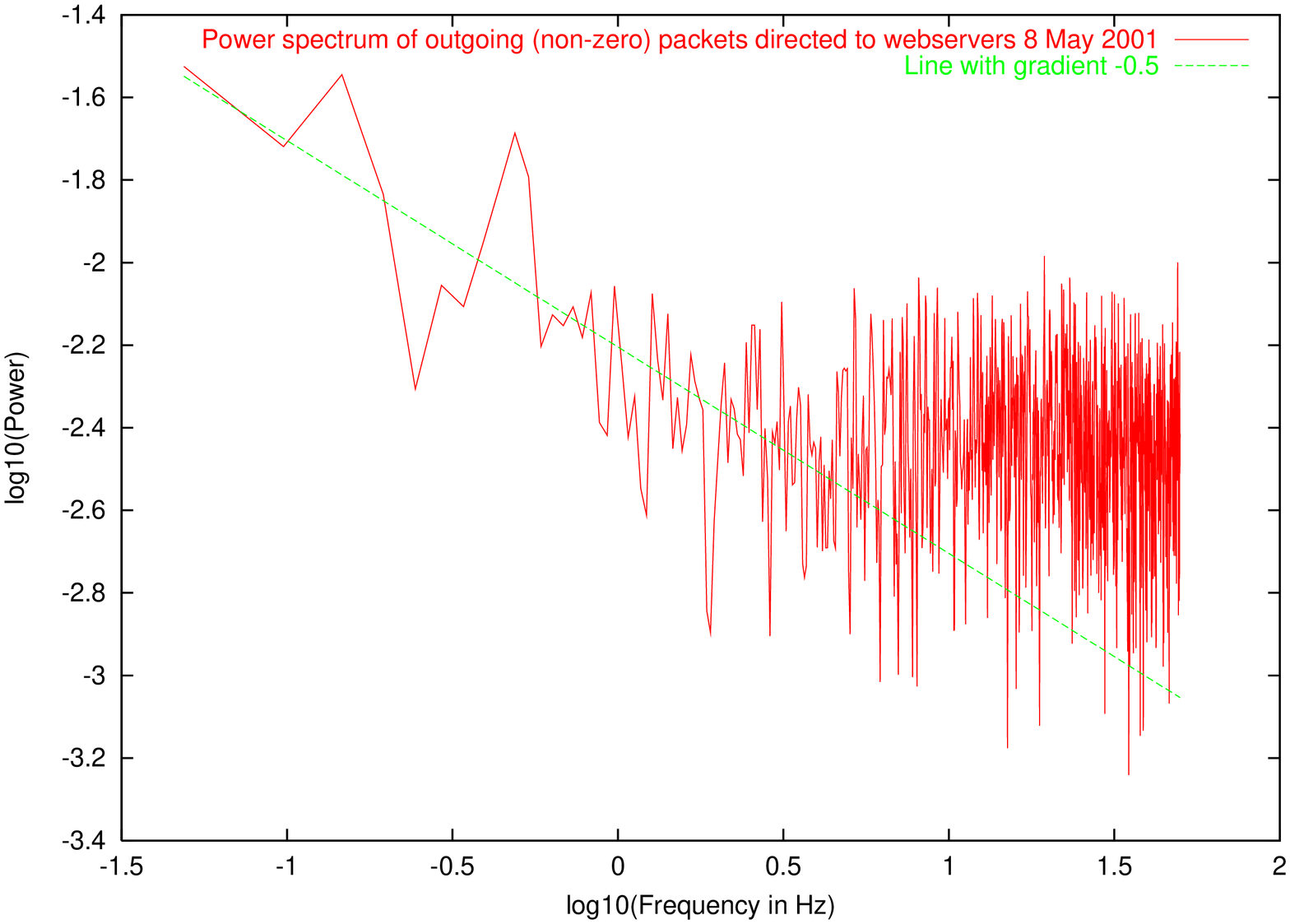}
\end{center}
\caption{\small Probability density function of the inter-arrival times
  for outgoing
  (non-zero) packets to web servers from the inside of DOC, i.e. requests for
  web pages. The power spectrum was taken on the 8 May 2001}
\label{fig:in-out}
\end{figure}

The last set of plots (fig. \ref{fig:in-out}) for the core router shows
the traffic created by 
users inside the department  that can be 
interpreted as requests for documents from an external server. Note
that the slope of the power spectrum is similar to that 
of the overall outgoing traffic, which seems sensible as the
web-traffic is a dominant part of the outgoing traffic. In contrast
the slope of the IIH is very different. This is not caused by the
neglect of small packets as they tend to contribute to the
left-most parts of histograms. 

So, the most interesting observation is that the incoming and outgoing
traffic for the internal web-servers appears to be uncorrelated. However,
the inter-arrival time distribution still follows a power law. We need to
check this result by, for instance, looking at log files of the
web-server. 

\subsection{Individual Node Traffic}
\label{node}
In this section we analyse traffic that we have monitored 
to and from a single
node in the student labs. 
\begin{figure}[tbp]
\hfill
\begin{center}
\includegraphics[height=6.5cm,angle=270]{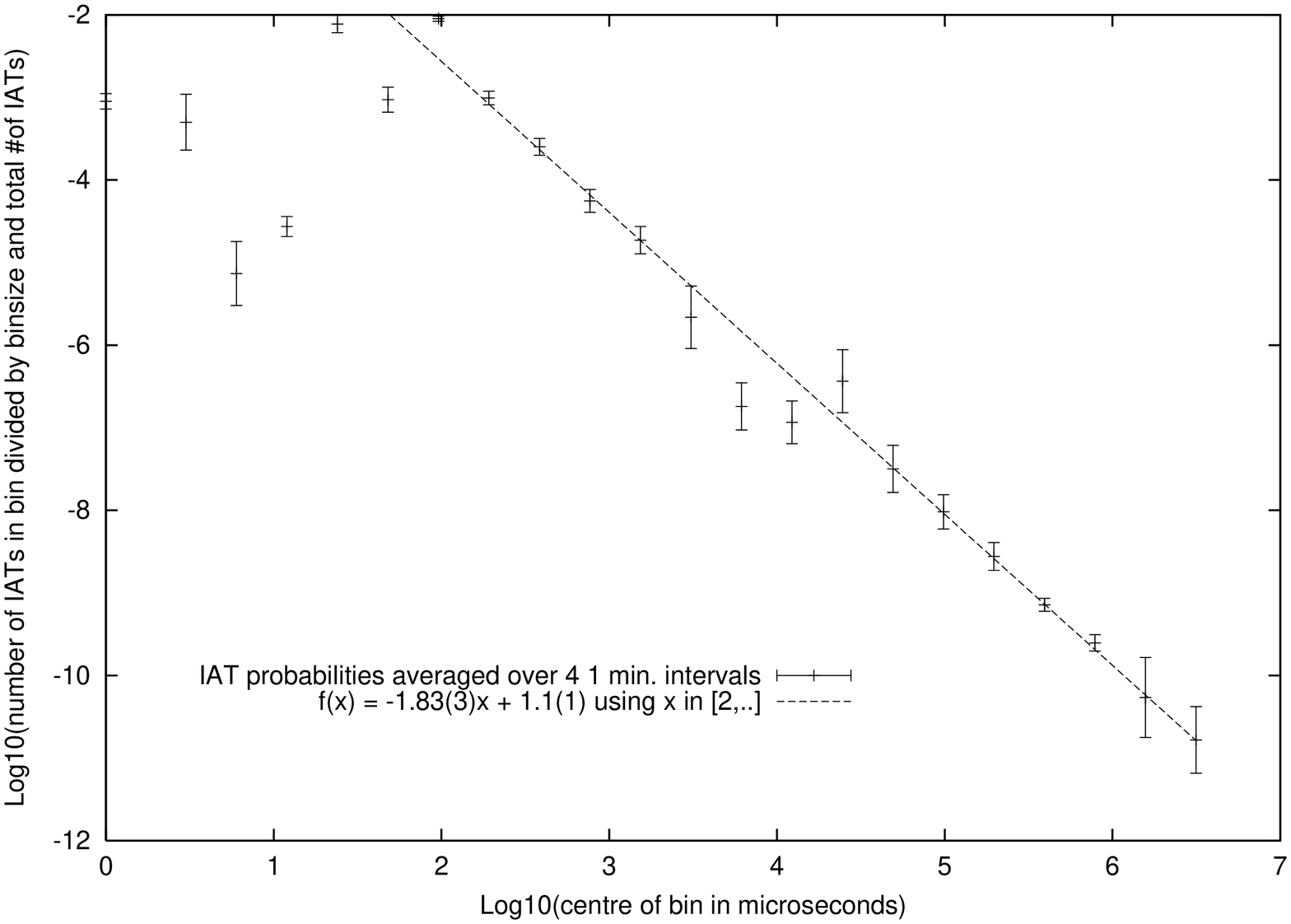}
\includegraphics[height=6.5cm,angle=270]{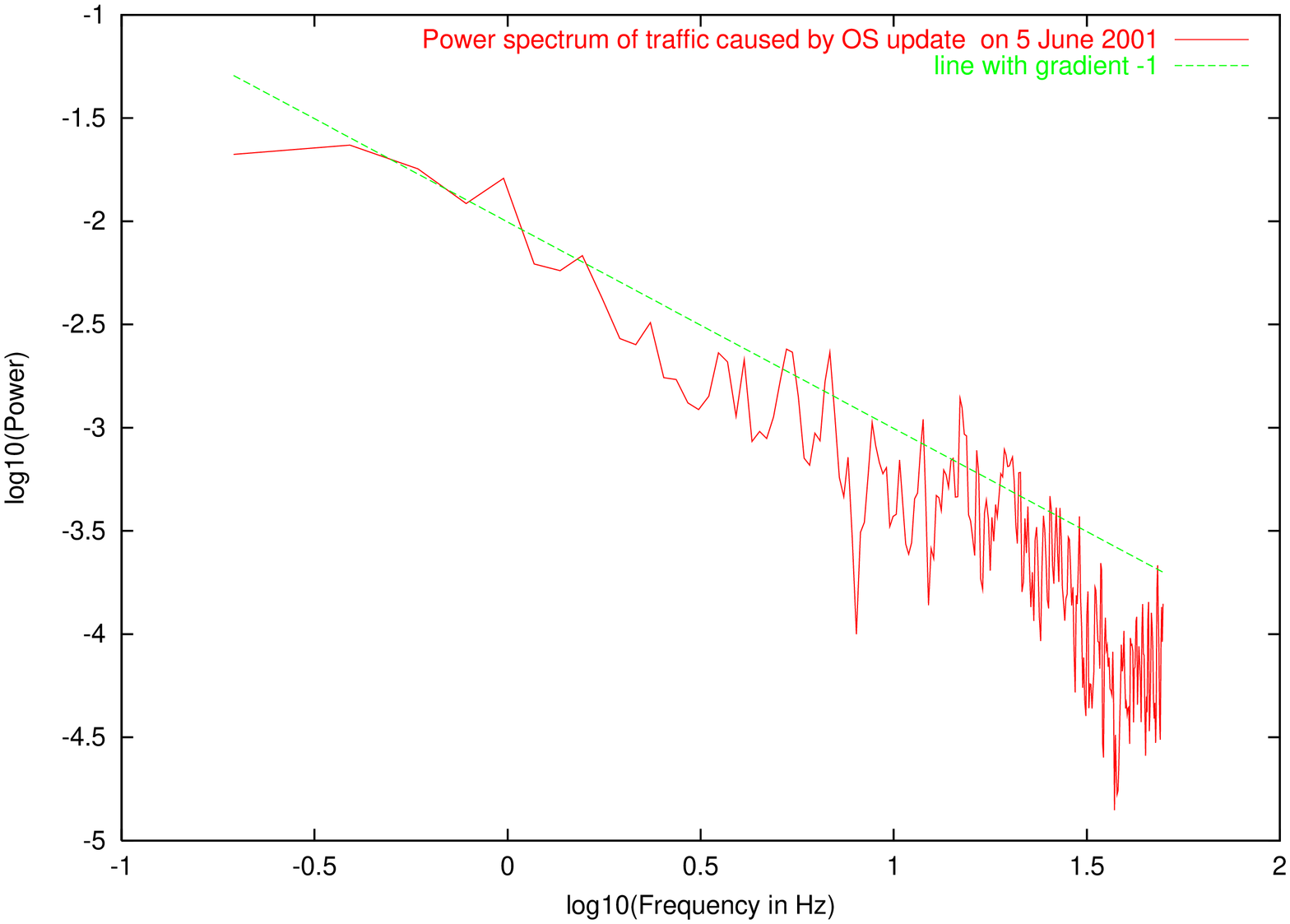}
\end{center}
\caption{\small Individual node traffic during over night update on 4
  and 5 June 2001 at 6.01-2am and 7.00-1am. This plot includes all
  packets. The power spectrum plot on the left is only for the 5 June.}
\label{fig:single-node-update}
\end{figure}

\begin{figure}[tbp]
\hfill
\begin{center}
\includegraphics[height=6.5cm,angle=270]{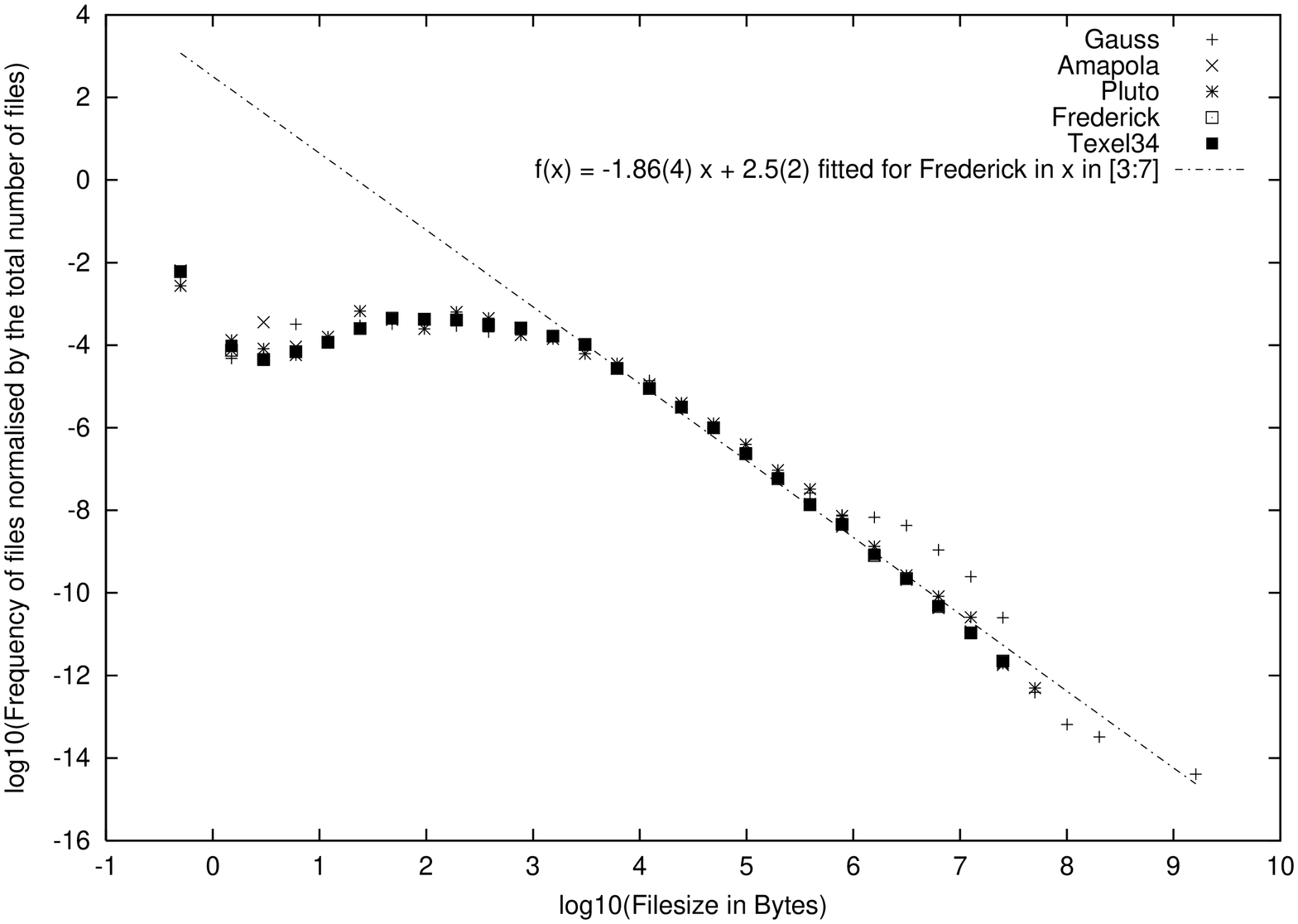}
\end{center}
\caption{\small File size distribution of 4 Linux machines and one
  Solaris Sparc machine (Pluto).}
\label{fig:unix}
\end{figure}

Every morning the OS on the LINUX machines in the student labs
gets updated by automatic scripts. These scripts update parts of the OS of the
target machine by pushing new packages (\texttt{rpm}s) to all LINUX
machines in the department. When we analysed the inter arrival
time distribution of all packets during that time we found that the
distribution still follows a power law
(fig. \ref{fig:single-node-update}). One reason for this could be 
the well-established fact that the distribution of file sizes in UNIX
systems follows a power law
\cite{UFSD}. Measurements taken for the node under investigation are
shown in figure (\ref{fig:unix}) \footnote{The graphs taken in figure
  (\ref{fig:unix}) show that this is true for the machine we
  monitored. The results are compiled with a slightly modified script
  from  \cite{UFSD}.}. On a
local homogeneous 
network the transmission speed  is constant, say 100 Mbps or
12.5 Mbytes/sec. Therefore we can measure the size of a file in units
of time: 1 Byte corresponds to roughly $80 \times 10 ^{-9}$ seconds. 
Assuming a program
pushes a large number of files over a local network as fast as
possible and the file size distribution is heavy-tailed, one would expect
the delivery  time of files to be following a heavy-tailed
distribution also. This still seems to be reflected in the
inter-arrival times
of frames, where a file corresponds to a number of frames, as the
maximum size of an Ethernet frame is about 1500 
Bytes. Also, the distribution in figure (\ref{fig:unix}) is static
in the sense that all files of the OS are included no matter whether
they are ever used or not. This is something to keep in mind when
web-servers are investigated: The distribution of the set of files
comprising the web-server may be very different to that of files requested
from the web-server. The total set might well include a large
number of unpopular or ``dead'' files.

We checked that the file size distributions on a number of our
UNIX/LINUX machines that include the  node monitored do follow a
power law  extremely well. Most surprisingly, we found that
the distribution for Solaris seems to follow exactly the same pattern
as that of a LINUX machine. 
Still we fail to see how the distribution of the delivery times of
files can produce the picture we see for the inter-arrival times of
frames on the Ethernet. Also, the slopes of distribution are
different. However this may provide a clue to the understanding of
the cause of the distribution. An interesting experiment would be to
monitor the transmission of a collection of files that is known to
have a geometric size distribution. In this case one would expect the
delivery time distribution to be Poisson.

\begin{figure}[tbp]
\hfill
\begin{center}
\includegraphics[height=6.5cm,angle=270]{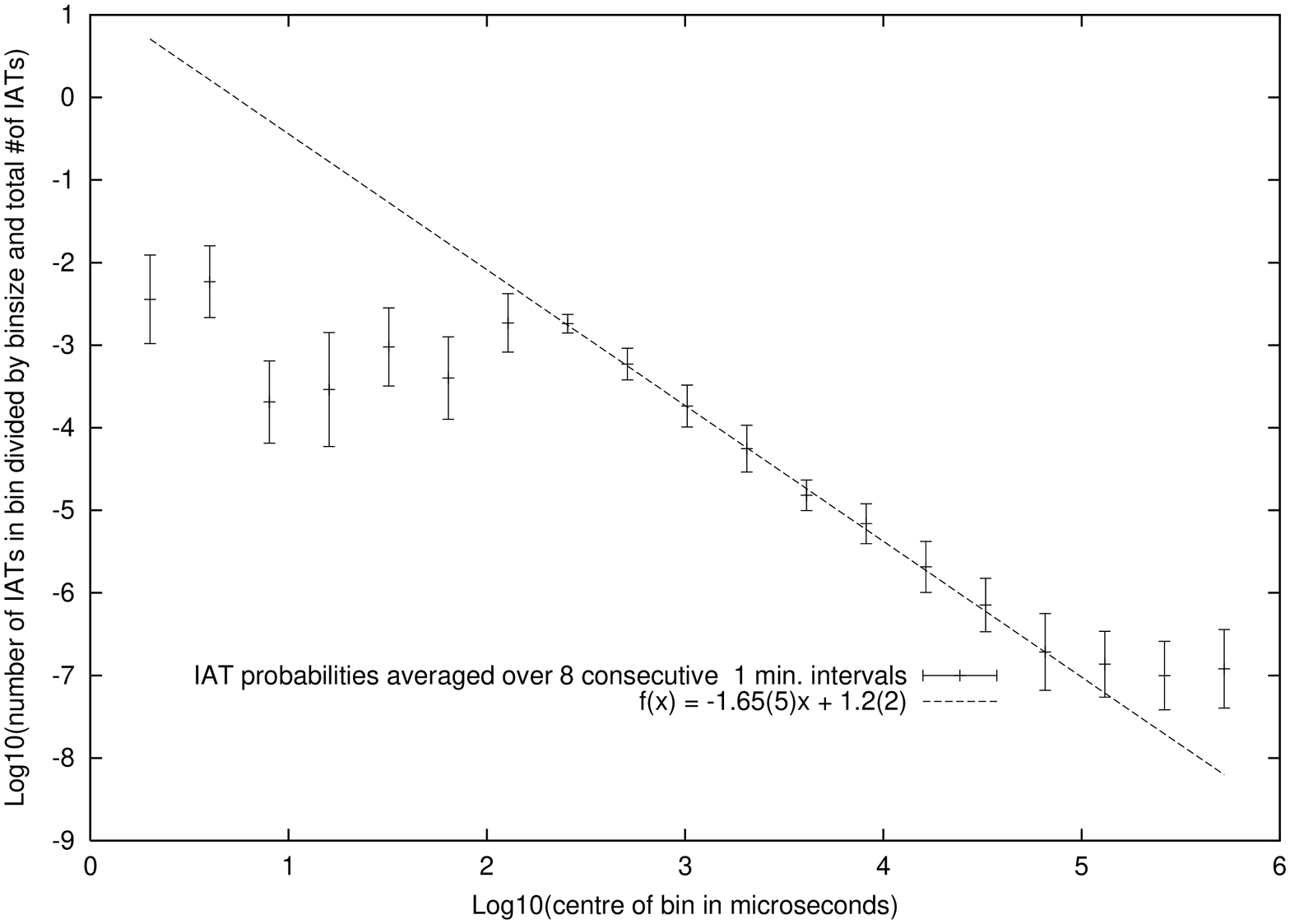}
\includegraphics[height=6.5cm,angle=270]{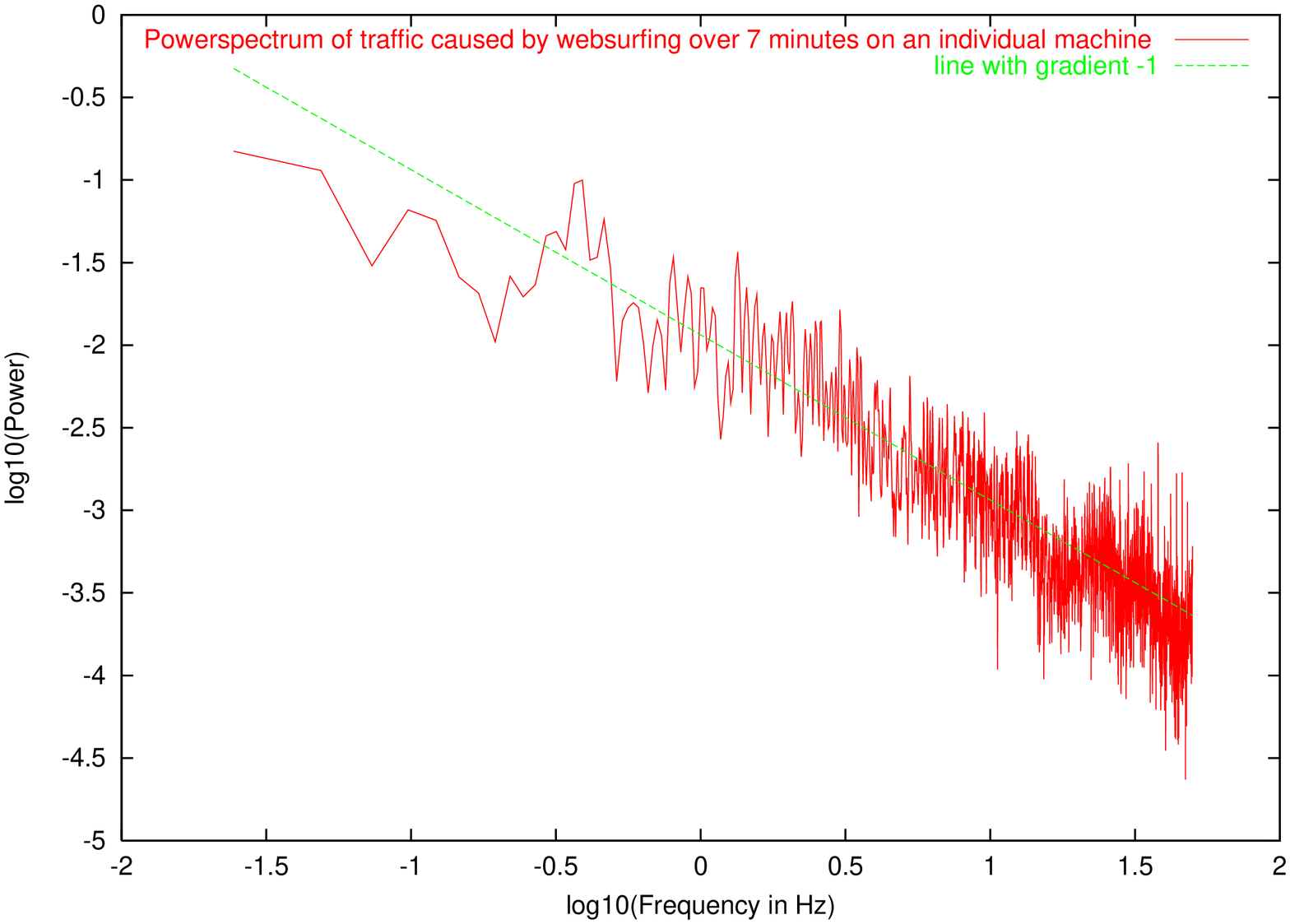}
\end{center}
\caption{\small Individual node traffic during the day whilst the
  computer was used for web-surfing. The measurements were taken
  between 1.48pm and 1.55pm on 4 June 2001. The plot shows  the
  inter arrival time distribution for all frames received and sent
  during that period and its power spectrum. }
\label{fig:single-node}
\end{figure}

\begin{figure}[tbp]
\hfill
\begin{center}
\includegraphics[height=6.5cm,angle=270]{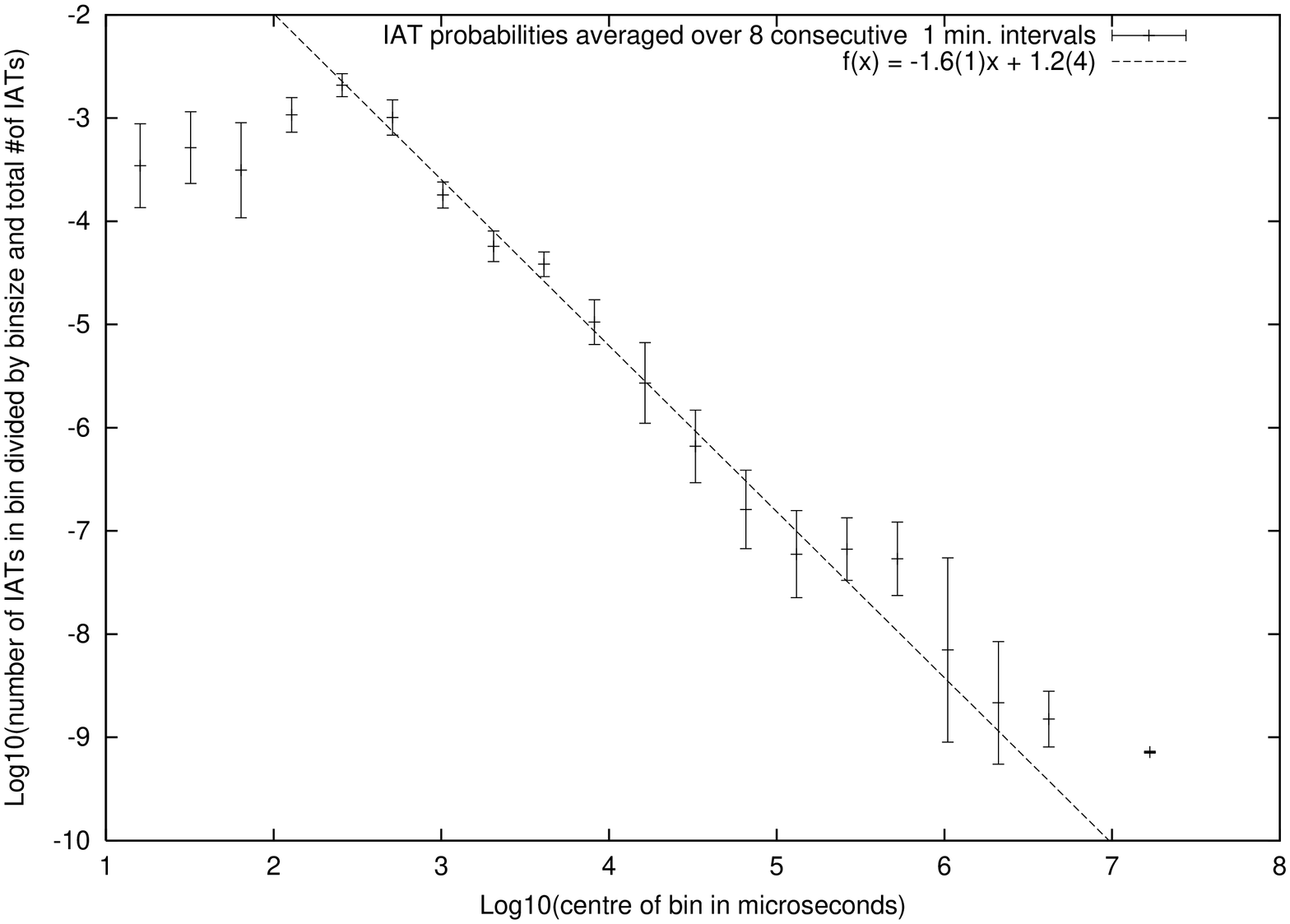}
\includegraphics[height=6.5cm,angle=270]{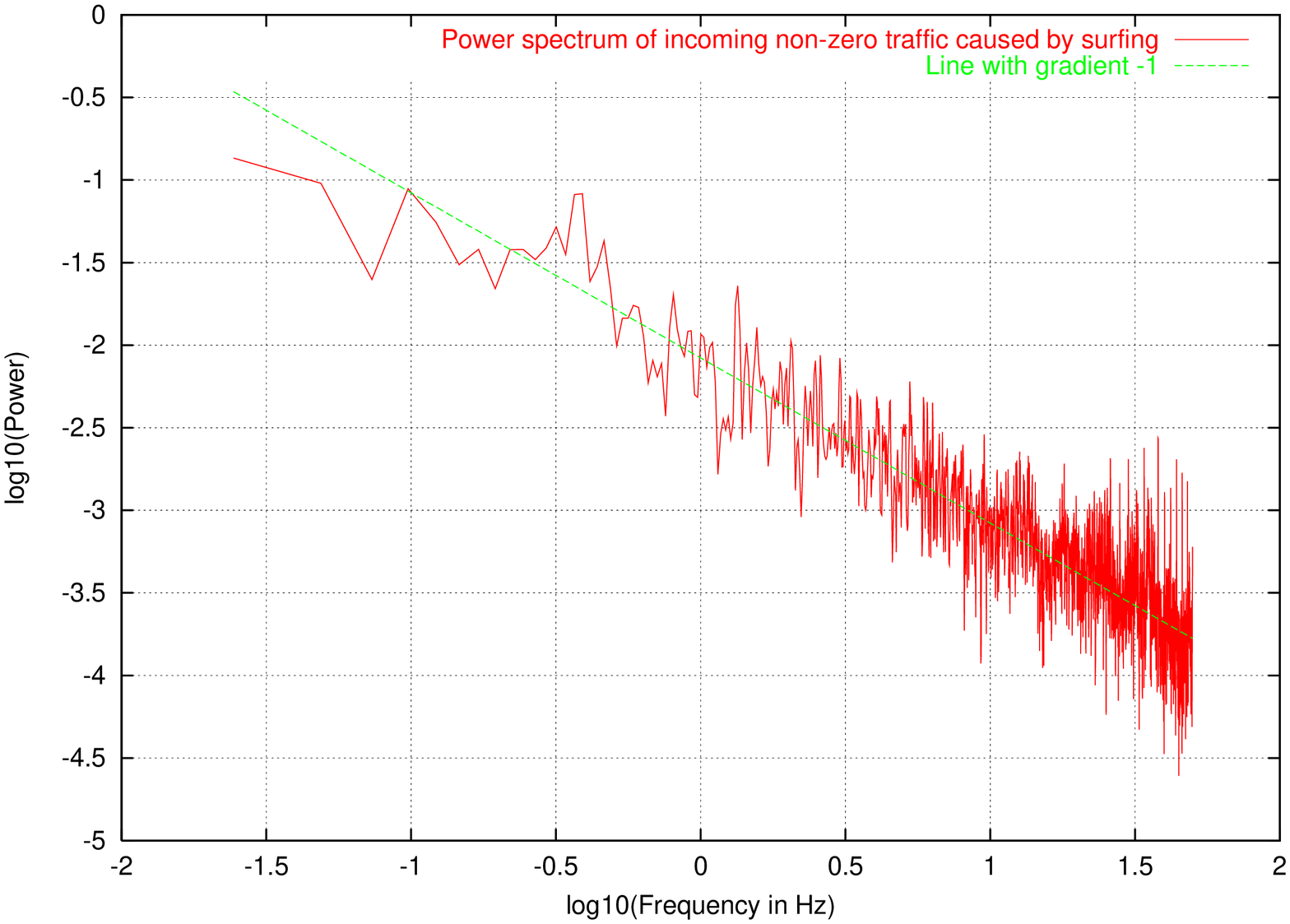}
\end{center}
\caption{\small Individual node traffic during the day whilst the
  computer was used for web-surfing. The measurements were taken
  between 1.48pm and 1.55pm on 4 June 2001. The plot only shows the distribution for incoming
  replies of non-zero frames and their power spectrum. }
\label{fig:single-node-wwwin}
\end{figure}

\begin{figure}[tbp]
\hfill
\begin{center}
\includegraphics[height=6.5cm,angle=270]{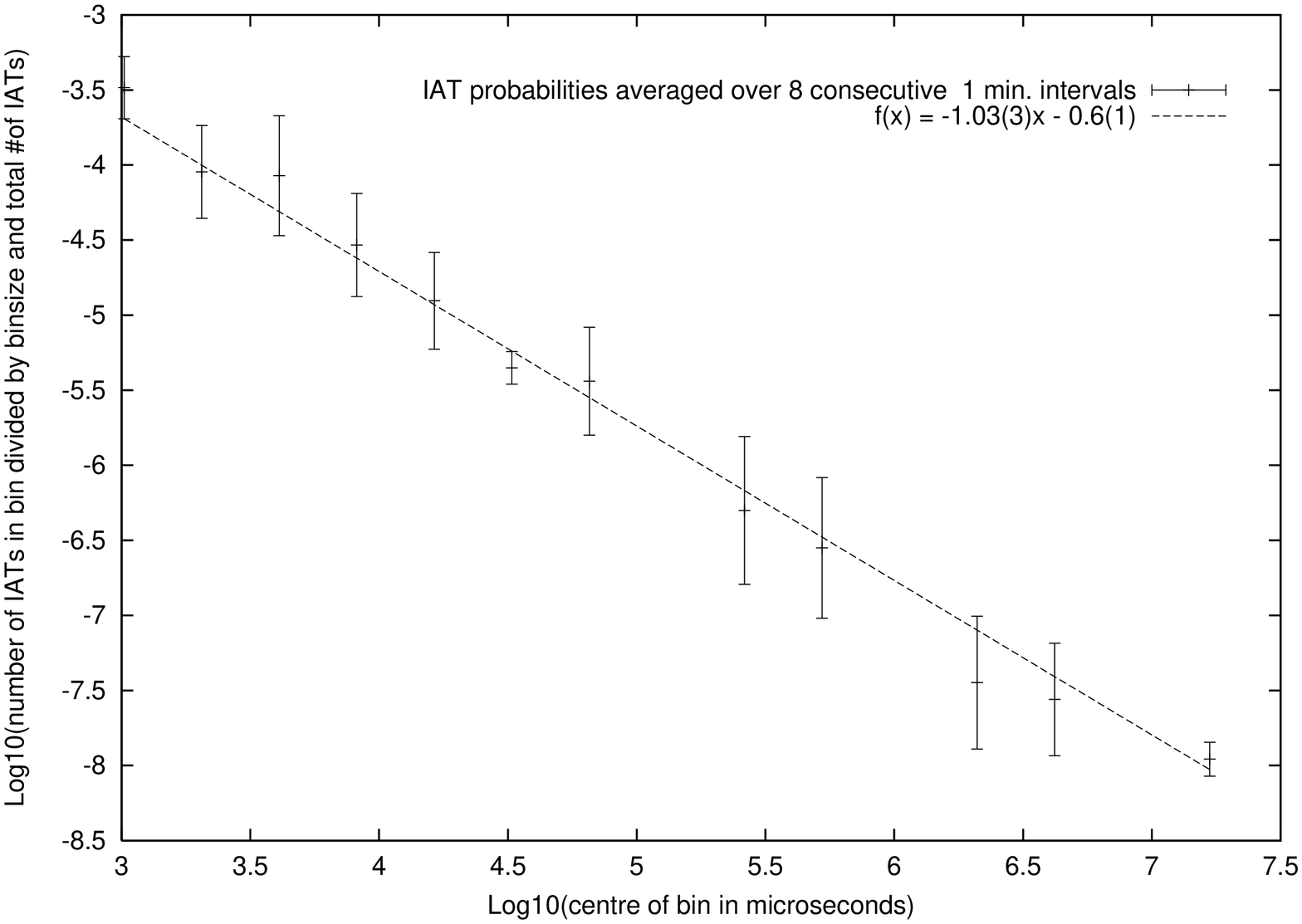}
\includegraphics[height=6.5cm,angle=270]{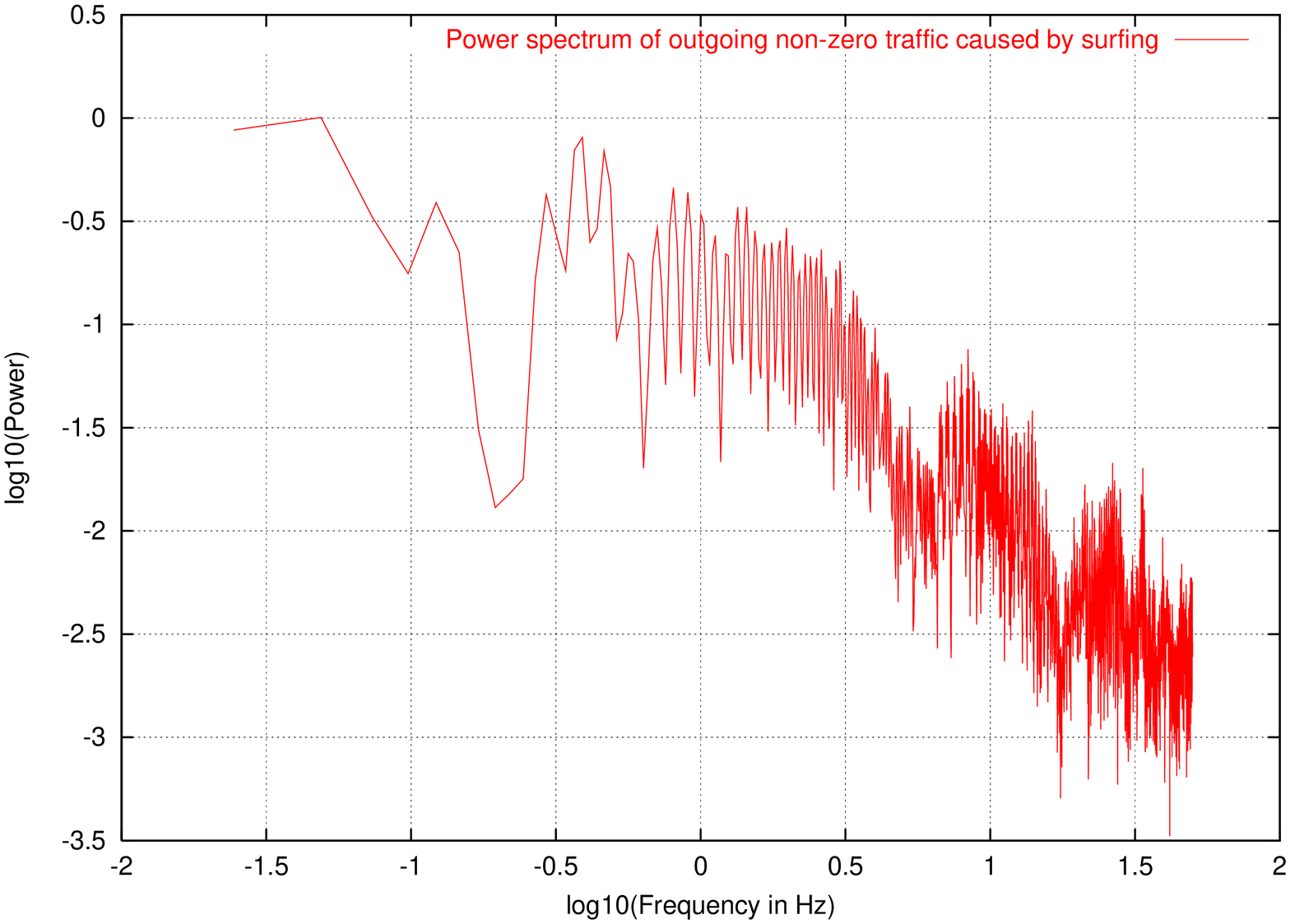}
\end{center}
\caption{\small Individual node traffic during the day whilst the
  computer was used for web-surfing. The measurements were taken
  between 1.48pm and 1.55pm on 4 June 2001. The plot shows the behaviour
  of outgoing web requests.}
\label{fig:single-node-wwwout}
\end{figure}

We also looked at the traffic behaviour caused by web-surfing
(fig. \ref{fig:single-node} - \ref{fig:single-node-wwwout}). Our 
criteria for web-related traffic is the same as in the previous
section. First we had a look at all frames sent to and received from the
node, i.e. not just the web-related traffic. Both the histogram and
the power spectrum suggest power 
laws. 
So the situation for traffic of a single type is different to that of
the core router where the entire traffic caused a power spectrum
without power law behaviour. Also, the slopes of the fitted lines do
not coincide with those fitted for the core router. This seems to
indicate a non-trivial aggregation process of the traffic, if one
assumes that internal nodes are fairly uniform\footnote{Of course the
  data for the internal web-servers already seems to suggest that nodes
  are not necessarily equal.}. 

For the received traffic,  we again
filtered out the zero packets (fig. \ref{fig:single-node-wwwin}). The
power spectrum of the incoming 
traffic appears to follow a power law describing $1/f$ noise. The
power spectrum seems to suggest a very strong long range correlation as
the exponent is very close to 1. 
From the relationship between the power spectrum and the
autocorrelation function  it can be shown that, a
gradient of the power spectrum slightly smaller than 1 implies
long-range dependency in the time series. 

The outgoing traffic looks less exciting but still seems to follow
power laws (fig. \ref{fig:single-node-wwwout}). In fact the observed
exponent is close to that estimated 
in aggregate traffic at the core router
(fig. \ref{fig:in-out}). However the exponent of the IIH is very
different, which may be related to the 
problem that this plot is 
based on a few hundred frames only in the observed time
period. Therefore one should  be careful in generalising  the
results. 

Earlier research into the behaviour of the inter-arrival time
distribution of modem connections made to an ISP has shown that the
distribution is 
essentially exponential, i.e. the traffic is Poisson
\cite{FGWK}. 
Filtering for packets that are non-zero and heading
towards a server gives us every frame related to a ``GET/POST''
command, when we look at the outgoing web-requests. So, we do not
observe the time between mouse-clicks users make when surfing with
their web 
browser, as 
e.g. in-lined images and cascaded style sheets will spoil our data by
triggering extra requests.  Still this data should a close
approximation of the time between successive mouse-clicks
Assuming that the web-surfing behaviour of our users is no different to
those dialling in via an ISP, this emphasises the need to model network
behaviour differently at different levels. While the connection
creation my 
follow an exponential behaviour \cite{FGWK} the actual data transfer
during the 
connection does not.

\subsection{Traffic generated by an MMPP}
\label{mmpp}
There have been various papers claiming a successful modelling of
network traffic using Markov modulated Poisson processes
\cite{Vaton}. This work was inspired by earlier work that successfully
showed that MMPPs can exhibit a Hurst coefficient for a limited range
of time scales \cite{Robert}. Also, there are ways to fit given data to an
MMPP using various methods \cite{MMPP-fit}.

\begin{figure}[tbp]
\hfill
\begin{center}
\includegraphics[height=6.5cm,angle=270]{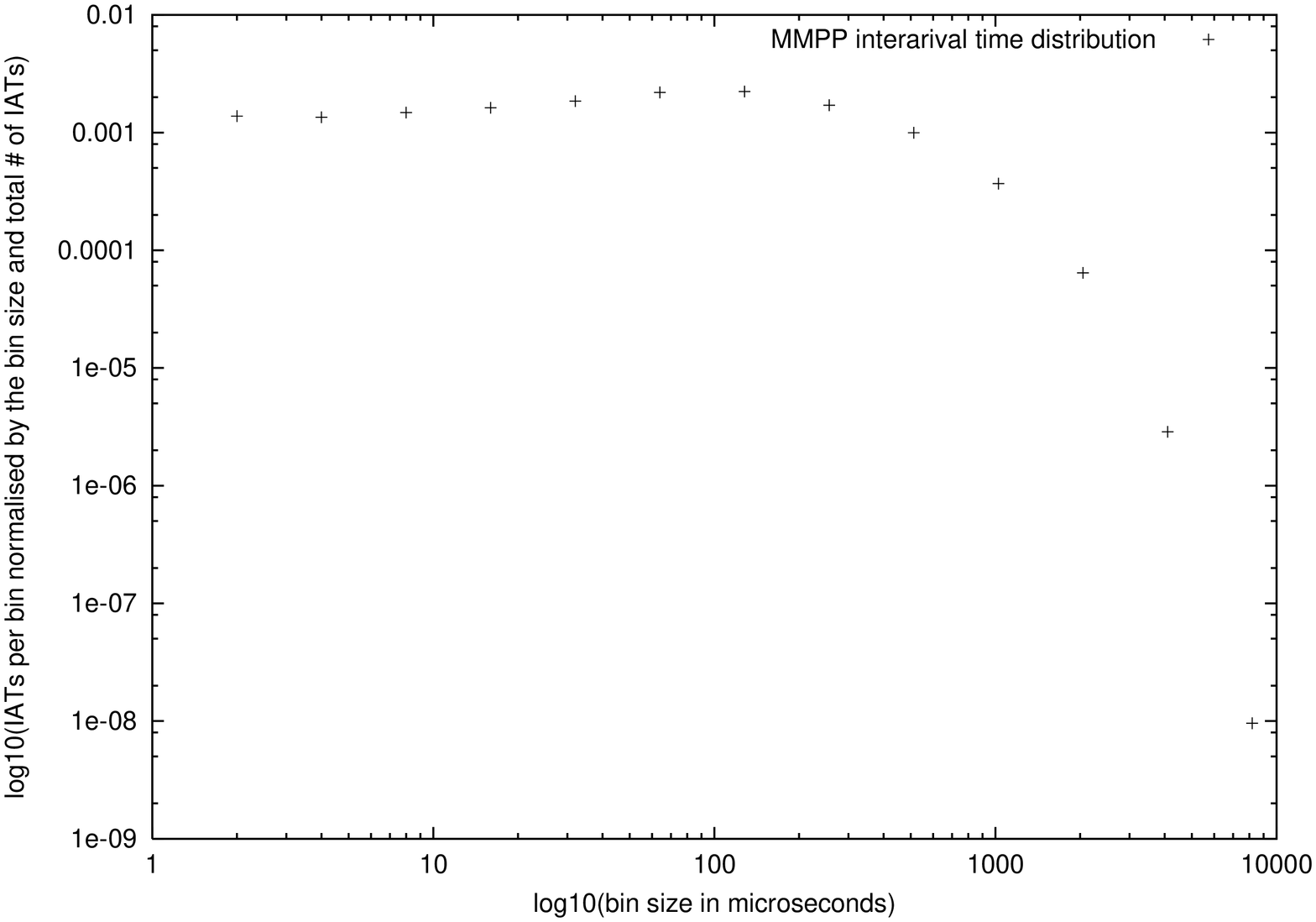}
\includegraphics[height=6.5cm,angle=270]{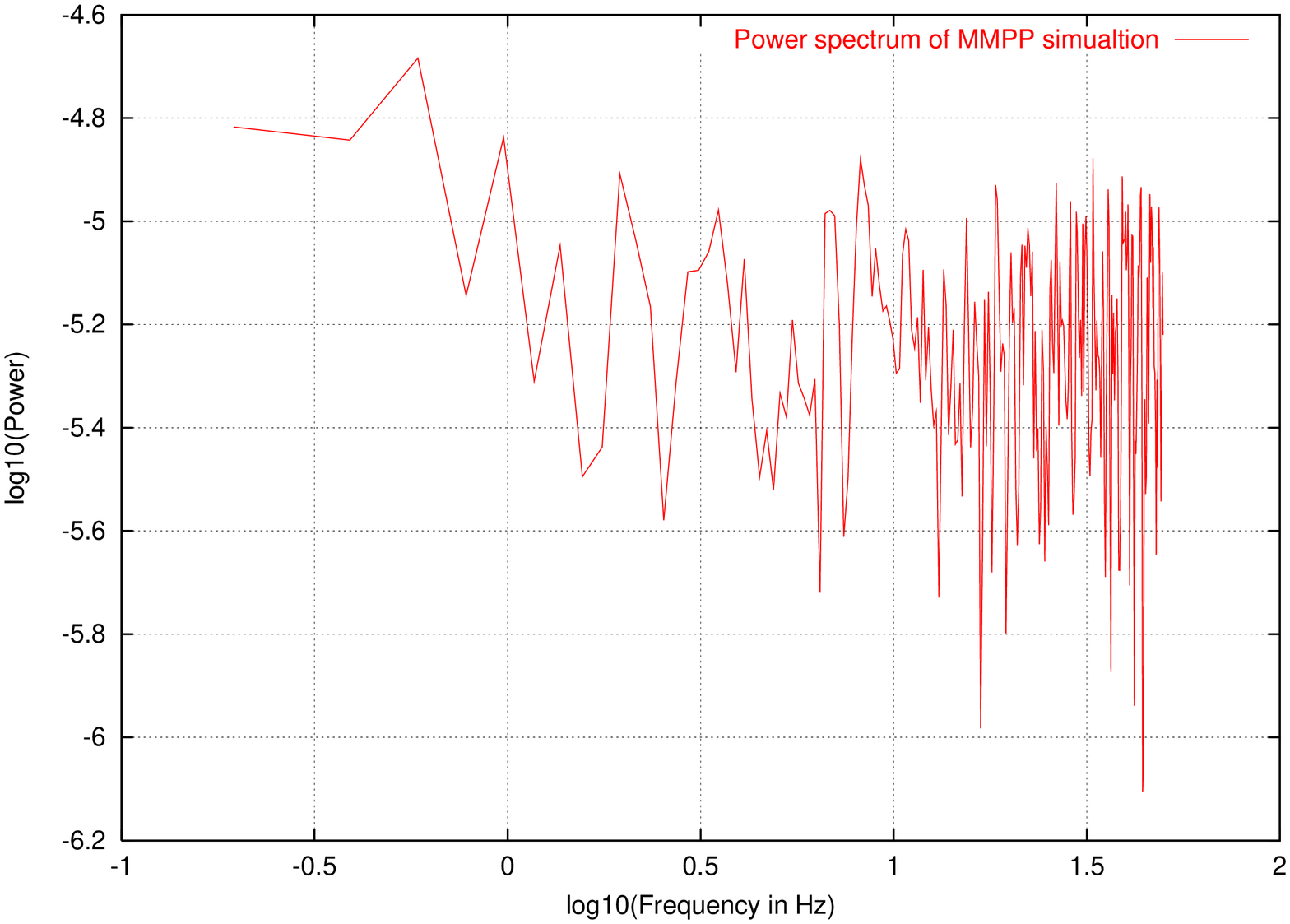}
\end{center}
\caption{\small Example histogram and power spectrum for a simulated  2 state
  MMPP}
\label{fig:mmpp}
\end{figure}

We have begun preliminary studies of using MMPPs to model the network
traffic we have observed. We ran simulations of a 2 state MMPP with
realistic choices for the 
arrival rates and packet sizes (though their distribution is not
bi-modal). The resulting histograms differ clearly from the ones we
have shown for the real traffic. None of the examples we looked at
showed a suitable IIH. But one should keep in mind that this hope is
very unrealistic as the IIH is simply the combination of two Poisson
processes, which will not exhibit a power law. However, the power
spectrum can be 
made to look realistic (\ref{fig:mmpp}). This casts doubts on the
ability to use 
MMPPs for realistic modelling of network traffic, as it should
certainly capture the features shown for the inter-arrival time
distribution. It 
may be possible to achieve better results with the use of more
states. The interpretation of the number of states may, however, be
difficult. Ideally, one would have hoped to be able to have a
correspondence between states and types of traffic. However, the
results from the filtered web-traffic show that even for one
particular traffic type, the inter arrival time distribution is more
complicated than a simple exponential. Also, more subtle extensions to
the idea of MMPPs may be more successful in capturing the statistical
features exhibited by the network traffic \cite{megan}. One possibility
is that MMCPPs, which include (compound) arrivals and
departures  of geometric size, show a more realistic
behaviour. 
Any MMPP must give exponential tails in the inter-arrival time density
function, being a (varying) mixture of Poisson processes. Consequently
there is no chance of representing a polynomial tail. However, a given
arrival process might be approximated ``up to the polynomial tail'',
the point at which the approximation becomes poor pushed sufficiently
to the right by suitable parameter selection. 
This work is still very much in progress.

\section{Conclusions, work in progress and future directions}
\label{sec:conclusion}
We have shown that different types of network traffic show power law
behaviour in both the IIH and their power spectra. The correlation of
the traffic as indicated by the power spectrum can be very low
(outgoing web traffic for a single node) to very high (incoming
traffic for a single node). 

Recently, there have been claims that self-similarity seen in network
traffic are caused by the CSMA/CD algorithm used in shared
Ethernets \cite{collision-fractal}. Since our network is full-duplex
the automatic update of the Linux machines should not experience any
collisions. Still the power spectrum seems to indicate that the
traffic has some long-range dependence. We need to investigate further
whether this could be caused by some hidden parts of the network
topology. Otherwise the results would suggest that self-similarity of
network traffic can occur without collisions. The automatic traffic
may be a good starting point for a further investigation as we do not
have the added problem of indeterminate user behaviour, but rather the
demand is caused by a simple script. It remains to be seen though how
these results could be used to explain the behaviour of router as they
tend to handle multiple sources of traffic.

Soon, the departmental connection will be upgraded to a 1Gbps link. We
would like to extend our monitoring capabilities to that link. One
problem we will face is that the 1 $\mu$sec resolution of
\verb|tcpdump|  running on a PC will no longer be sufficient to distinguish
between packet arrivals when the link is fully utilised. However,
other reports seem confident that \texttt{tcpdump} can still be useful for
monitoring faster links \cite{GB}. 

Another aim is to automate the way power spectra
and IIHs are generated to 
enable us to observe the results for a long period of time and see
whether we can use the results to generalise the traffic behaviour we
have observed so far. 

Our initial hope had been to use the results of the analysis to fit a
fairly simple MMPP. As this has proved elusive, we will investigate
whether  there is a time regime in which MMPPs or MMCPPs are
adequate to model the network traffic. If this is the case, it may be possible
to treat the remaining traffic on a longer time scale causing  the network
traffic to change from one phase to another, for example. We will have to
determine that time scale. 
In such a two-time-scale model, the main issues will be
\begin{enumerate}
\item to find the equilibrium between events on the longer time scale
  (e.g. router rebooting)
\item to find the time constant for a new equilibrium to be reached  after
  each type of event
\item to investigate the transient behaviour immediately after each
  type of event. 
\end{enumerate}

A possible interpretation of  $1/f$ noise is the notion of self-organised
criticality (SOC) \cite{BTW,J}. The word criticality is borrowed from
physics where a critical state of a system is related to an infinite
correlation length and the system going through a phase
transition. There are many areas in science where $1/f$ has been seen:
see for instance \cite{J} for a good overview of the topic. We would
like to investigate whether the systems we have been  monitoring are really
exhibiting $1/f$ noise and if this can tell us, similar to the noise
observed in heart beats of humans, whether the network is in a
``healthy'' or ``sick'' state.

Recently the analysis of time series in the context of financial data
has seen a great deal of interest. We need to investigate how we can
utilise those results for our purpose, as the behaviour
of the data and aims of its analysis seem similar \cite{FB}. The same
is true for models of car 
traffic on roads. There have been great advances in the modelling and
prediction of car traffic using, for  instance, cellular
automata. Models capture the critical behaviour of real car traffic,
like phantom 
traffic jams \cite{traffic}. Indeed the models are good enough to be
used for the 
prediction of city and motorway traffic. So, how does the traffic seen
in Etherland correspond to real car traffic? Well, in the state of
Full-Duplex the rules on a 100Mbps line are simple and resemble that
of a winding 
country road:  
\begin{enumerate}
\item There is no overtaking.
\item There is a universal speed limit, and all participants drive up
  to the limit (100Mbps $\approx 12.5$ MBps) . 
\item The participants in the traffic are safety conscious and keep a
  distance to the packet ahead, the interframe gap (IFG). This is the
  length of 12 Bytes or approximately 1$\mu$sec.
\end{enumerate}
There are however some significant differences:
\begin{enumerate}
\item Associating the length of cars and lorries with that of packet
  sizes in Bytes, one finds that the ratio of small to big packets is
  about 23, so Etherland lorries are huge. For road traffic this
  ratio will be close to between 4 or 8.
\item Assuming we monitor the traffic as it passes through a point in
  the network, it takes a small packet about 5$\mu$sec to go past. In
  real traffic assuming a speed limit of 80km/h (50mph) a 5m long car
  takes about 1/4 second to cover its own length. This difference has
  an interesting effect on the observed data. To observe as many
  events in road traffic as one does in, say, 5 minutes of network
  traffic one has to observe road traffic for several hours.
\item Though there are conservation laws for cars, data can be
  annihilated without trace at any time in any place of the system. In
  models of car traffic one can make assumptions that cars that are being
  driven to work in the morning are most likely to return to their
  garages in the evening. For data traffic this is  not necessarily the
  case.
\item Another curious feature, if one wants to associate a ``car'' with a
  data packet, is that data packets never see other traffic whilst on
  the road; only in buffers is it possible for them to ``see'' the
  surrounding traffic. 
\end{enumerate}
For a recent publication dealing with a model for TCP traffic using
cellular automata, see \cite{HBKSS}.

\section*{Acknowledgements}
The authors would like to thank the Computer Support Group and in
particular Stuart McGregor and David Wragg in helping with the data
capture. We would also like to thank Maya Paczuski  and J\"orn
Davidsen of the Maths Department of IC for fruitful discussions on
self-similarity and criticality, and  Will Knottenbelt and David
Thornley for stimulating conversations.

The data used for the investigation can be made available on request,
though we will have to anonymise it due to data protection
issues. Similarly, the scripts used to perform the actual data
monitoring are available.

The research was funded by EPSRC (grant QUAINT).

\end{document}